\documentclass{acta}

\usepackage{amsmath}
\usepackage{amssymb}
\usepackage{rotating}
\usepackage[hidelinks]{hyperref} 
\usepackage{bm}

\usepackage{xcolor}

\newcommand\tE{t_{\rm E}}
\newcommand\thetaE{\theta_{\rm E}}

\newcommand\pirel{\pi_{\rm rel}}

\newcommand{\dummy}[1]{}

\begin{document}

\begin{Titlepage}

\Title{\textit{Eppur non si trovano}: Comments on the Primordial Black Hole Limits in the Galactic Halo}


\Author{Przemek Mr\'oz$^1$, Andrzej Udalski$^1$, Micha\l{} K. Szyma\'nski$^1$,\\ Igor Soszy\'nski$^1$, \L{}ukasz Wyrzykowski$^1$, Pawe\l{} Pietrukowicz$^1$,\\ Szymon Koz\l{}owski$^1$, Rados\l{}aw Poleski$^1$, Jan Skowron$^1$, Dorota Skowron$^1$, Krzysztof Ulaczyk$^{1,2}$, Mariusz Gromadzki$^1$, Krzysztof Rybicki$^1$,\\ Patryk Iwanek$^1$, Marcin Wrona$^{1,3}$ and Milena Ratajczak$^1$}
{$^1$Astronomical Observatory, University of Warsaw, Al. Ujazdowskie 4, 00-478 Warszawa, Poland\\
e-mail: pmroz@astrouw.edu.pl\\
$^2$Department of Physics, University of Warwick, Coventry CV4 7AL, UK\\
$^3$Department of Astrophysics and Planetary Sciences, Villanova University, 800 Lancaster Avenue, Villanova, PA 19085, USA
}

\Received{MM DD, YYYY}

\end{Titlepage}

\Abstract{
In a recent arXiv post, Hawkins \& Garc\'ia-Bellido raised doubts on the results of 20-yr long OGLE photometric monitoring, which did not find a large number of gravitational microlensing events in the direction of the Magellanic Clouds. These results implied that primordial black holes and other compact objects with masses from $10^{-8}$ to $10^3\,M_{\odot}$ cannot comprise a substantial fraction of the Milky Way dark matter halo. Unfortunately, the Hawkins \& Garc\'ia-Bellido post contained a number of scientific misrepresentations of our work. 
Here, we demonstrate that their arguments lack a solid basis or are simply incorrect.  As we show below,  {\it and yet they are not found} -- compact objects (including primordial black holes) in the dark halo of the Milky Way remain undetected, despite extensive searches.}
{Gravitational microlensing (672), Dark matter (353), Milky Way dark matter halo (1049), Primordial black holes (1292)}

\section{Introduction} \label{sec:intro}

Since the seminal paper by \citet{paczynski1986}, gravitational microlensing has been regarded as an important tool for studying dark matter in the Milky Way. If the entire Milky Way dark matter halo were made of compact objects, such as primordial black holes (PBHs), we would expect to observe a large number of gravitational microlensing events in the direction of the Magellanic Clouds or the M31 galaxy. Paczy\'nski's paper inspired the formation of the first sky variability surveys, such as MACHO, EROS, or OGLE. These projects aimed to observe millions of stars over extended periods of time to search for telltale light curve variations that are characteristic of gravitational microlensing events.

The culmination of these efforts was the study conducted by \citet{mroz2024a, mroz2024b}, which reported the results of searches for gravitational microlensing events in the light curves of nearly 80 million stars located in the Large Magellanic Cloud (LMC) collected from 2001 to 2020 during the third and fourth phases of the OGLE survey \citep{udalski2003,udalski2015}. The study identified only 16 microlensing events (13 of which were selected using automated detection procedures), far too few compared to the number expected if the entire Milky Way dark matter halo were composed of compact objects with masses ranging from $10^{-5}$ to $10^3\,M_{\odot}$. In addition, the follow-up paper by \citet{mroz2024d} did not find any short-timescale microlensing events in the Magellanic Clouds, placing strong limits on the frequency of planetary-mass ($10^{-8}\!-\!10^{-2}\,M_{\odot}$) PBHs. \citet{mroz2025} presented the results of searches for gravitational microlensing events toward the Small Magellanic Cloud (SMC).

Recently, this work was criticized in an arXiv post by \citet{hawkins2025}. In response, we refute their arguments point by point. The central line of arguments brought forward by \citet{hawkins2025} is connected to the discrepancy between the results obtained from the OGLE project and those published by the MACHO project in the early 2000s \citep{alcock2000b}. In this discussion, we will first compare the MACHO and OGLE data sets (Section~\ref{sec:macho}), present new reductions of the MACHO data (Section~\ref{sec:dia}), and then we will examine the arguments put forth in the \citet{hawkins2025} post (Sections~\ref{sec:selection} to \ref{sec:models}).

\section{MACHO Events}
\label{sec:macho}

The MACHO project used the 1.27 m telescope at Mount Stromlo Observatory, Australia, which was equipped with a camera that captured a field of view of $42' \times 42'$ simultaneously in two filters. The study by \citet{alcock2000b} analyzed the photometry of 11.9 million stars in the LMC collected during the first 5.7 years of observations from the MACHO project. They reported the discovery of 13 to 17 microlensing event candidates (depending on the adopted selection criteria) and measured the microlensing optical depth toward the LMC to be $\tau=1.2 ^{+0.4}_{-0.3} \times 10^{-7}$. These findings were interpreted as evidence of compact objects with a mass of $0.60^{+0.28}_{-0.20}\,M_{\odot}$ that comprise $21^{+10}_{-7}\%$ of the Milky Way dark matter halo \citep{alcock2000b}.

The MACHO results have never been replicated by any other experiment searching for microlensing events in the direction of the Magellanic Clouds. The EROS survey \citep{tisserand2007} reported the discovery of only one microlensing event candidate, using a sample of 7 million bright stars observed in the LMC and the SMC over a total of 6.7 years. This led them to calculate the 95\% upper limit on the microlensing optical depth of $\tau<0.36 \times 10^{-7}$. 

Meanwhile, \citet{wyrzykowski2009}, using \mbox{OGLE-II} observations of 11.8 million stars monitored for nearly four years, identified only two microlensing event candidates. However, one of those candidates (OGLE-LMC-02) turned out later to be a variable star \citep{mroz2024a}. Observations of about 35 million stars in the LMC carried out during the third phase of the OGLE survey (\mbox{OGLE-III}) from 2001 through 2009 revealed only two strong microlensing event candidates. This resulted in a microlensing optical depth of $\tau=(0.16 \pm 0.12)\times 10^{-7}$ \citep{wyrzykowski2011}.

The most recent OGLE study, which combined \mbox{OGLE-III} and \mbox{OGLE-IV} observations spanning from 2001 through 2020, found 13 microlensing events selected using objective criteria \citep{mroz2024a,mroz2024b}. The derived microlensing optical depth toward the LMC is $\tau=(0.121 \pm 0.037) \times 10^{-7}$, in line with previous findings from both the EROS and \mbox{OGLE-III} studies of microlensing events toward that galaxy.

The MACHO data were independently analyzed by \citet{belokurov2004} and \citet{evans2005} (see also \citealt{griest2005} and \citealt{evans2007}), who used neural networks to classify the MACHO light curves. They argued that only six candidate events reported by \citet{alcock2000b} could be reliably classified as microlensing events (MACHO-LMC-(01, 05, 06, 14, 21, 25)). Additionally, two objects were considered microlensing candidates (MACHO-LMC-(09, 18)), while the rest were deemed unlikely to be microlensing events. \citet{evans2005} estimated a much lower value of the microlensing optical depth toward the MACHO fields ($0.3 \times 10^{-7} \lesssim \tau \lesssim 0.5 \times 10^{-7}$) than \citet{alcock2000b}.

Since the 1990s, a great amount of progress has been made in developing photometric techniques for dense stellar fields and in understanding the variable star contamination that affects microlensing experiments. One notable difference between the MACHO and OGLE data sets lies in the different methods employed to extract photometry. The MACHO project used a modified version of DoPHOT \citep{schechter1993}, which performs point spread function (PSF) fitting on science images \citep{alcock1999}. In contrast, the OGLE project implemented the Difference Image Analysis \citep[DIA;][]{tomaney1996,alard1998} method \citep{wozniak2000}, which is much better suited for dense stellar fields, such as those found in the LMC. 

\begin{figure}[htb]
\centering
\includegraphics[width=\textwidth]{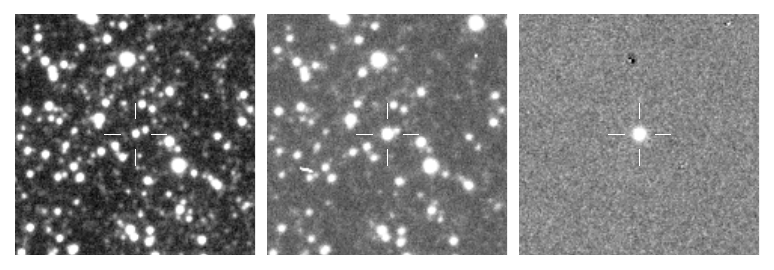}
\FigCap{OGLE observations of the event OGLE-LMC-12. Left panel: cutout from the reference image of the field LMC502.05. Middle panel: OGLE image from 2014 November 12. Right panel: subtracted image. The position of the event is marked with a white cross, North is up, and East is to the left. All images are $60'' \times 60''$. OGLE differential photometry is performed on subtracted images.}
\label{fig:dia}
\end{figure}

In DIA, several best-quality images of a given field (those with best seeing and lowest background) are stacked to create a reference image. This reference frame is then subtracted from the incoming images. The resulting subtracted image contains only those objects that have changed in brightness or position compared to the reference image, effectively removing all constant stars from the view, as shown in Fig.~\ref{fig:dia}.

Another important factor affecting the quality of photometry in dense stellar fields is seeing. According to \citet{alcock2001}, the typical seeing for MACHO template images (which were created using the highest-quality frames) ranges from 1.5 to $2.0''$, with a pixel scale of $0.63''$ per pixel. We downloaded the MACHO images from the online archive\footnote{https://macho.nci.org.au/} and found the median seeing in the $R$-band images to be $2.7''$. In contrast, the reference images from the \mbox{OGLE-III} and \mbox{OGLE-IV} surveys have a typical full width at half maximum (FWHM) of $0.9\!-\!1.0''$, and the median seeing on science images is $1.25\!-\!1.35''$ \citep{udalski2015}. The \mbox{OGLE-III} and \mbox{OGLE-IV} cameras have pixel sizes of $0.26''$. These factors, when combined, result in a significantly improved quality of the photometric measurements.

\begin{figure}[htb]
\centering
\includegraphics[width=\textwidth]{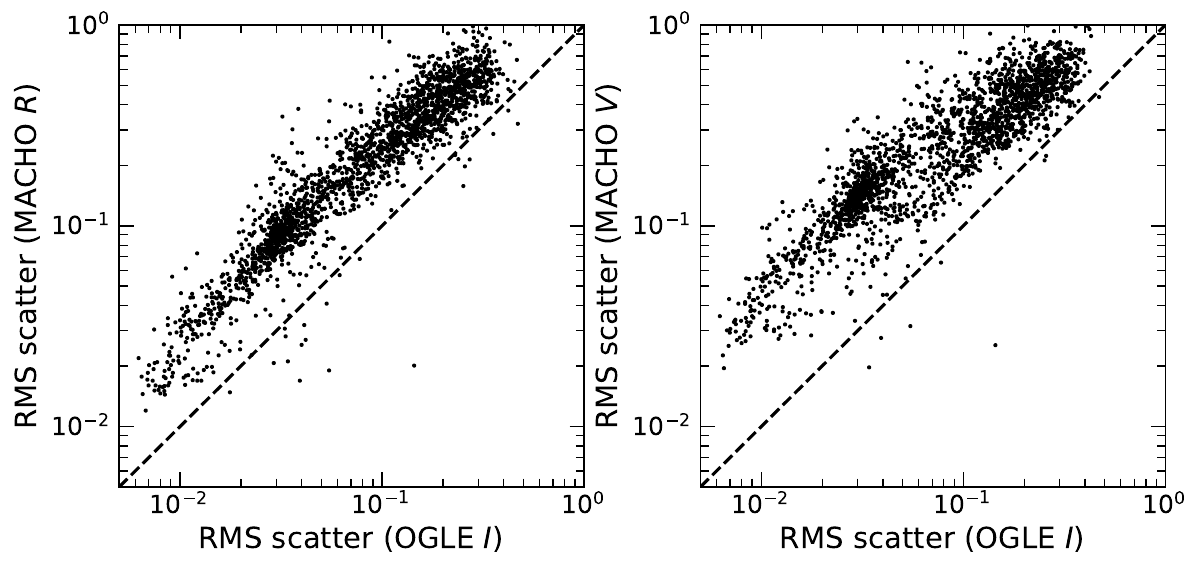}
\FigCap{Comparison between the rms scatter (in magnitudes) in the MACHO $R$-band (left panel) and $V$-band (right panel) light curves and the rms scatter in the OGLE light curves of same stars.}
\label{fig:rms}
\end{figure}

This improvement is illustrated in Fig.~\ref{fig:rms}, which shows the root mean square (rms) scatter for a sample of about 3000 stars located in the MACHO field 6.5967 and the OGLE field LMC502.22. The MACHO light curves (which were generated using MACHO PSF fitting pipeline) were downloaded from the online archive. The left and right panels present the rms scatter in the MACHO $R$ and $V$ bands, respectively, and the rms scatter in the OGLE light curves is calculated using $I$-band observations of the same stars. We used the same code to calculate the rms scatter in both the MACHO and OGLE data. On average, the rms scatter in the MACHO $R$-band ($V$-band) light curves is a factor of $2.45 \pm 0.02$ ($3.08 \pm 0.03$) larger than in the OGLE light curves. The discrepancy is even more pronounced for OGLE stars brighter than $I=19$\,mag for which the rms scatter in the MACHO $V$-band light curves is larger by a factor of $4.54 \pm 0.06$ than in the OGLE light curves.

The difference in the rms scatter between the MACHO and OGLE light curves explains why the MACHO data set is more susceptible to contamination from variable stars. Variability in the baseline may be buried in the noise, or certain light curve features may be washed out by the noise, making it more difficult to differentiate variable sources from genuine microlensing events. 

\begin{figure}[htb]
\centering
\includegraphics[width=\textwidth]{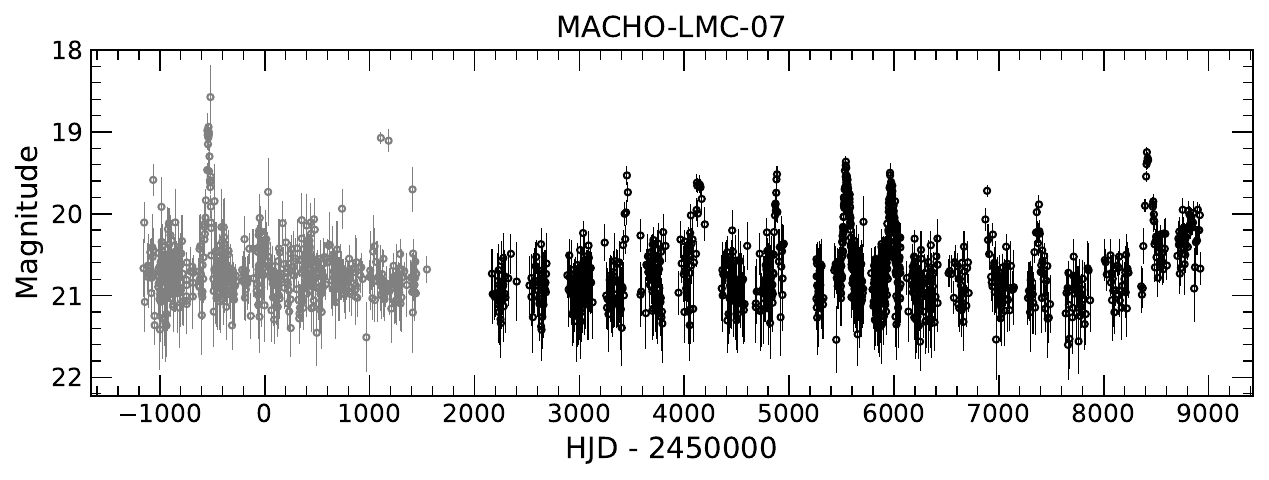}
\includegraphics[width=\textwidth]{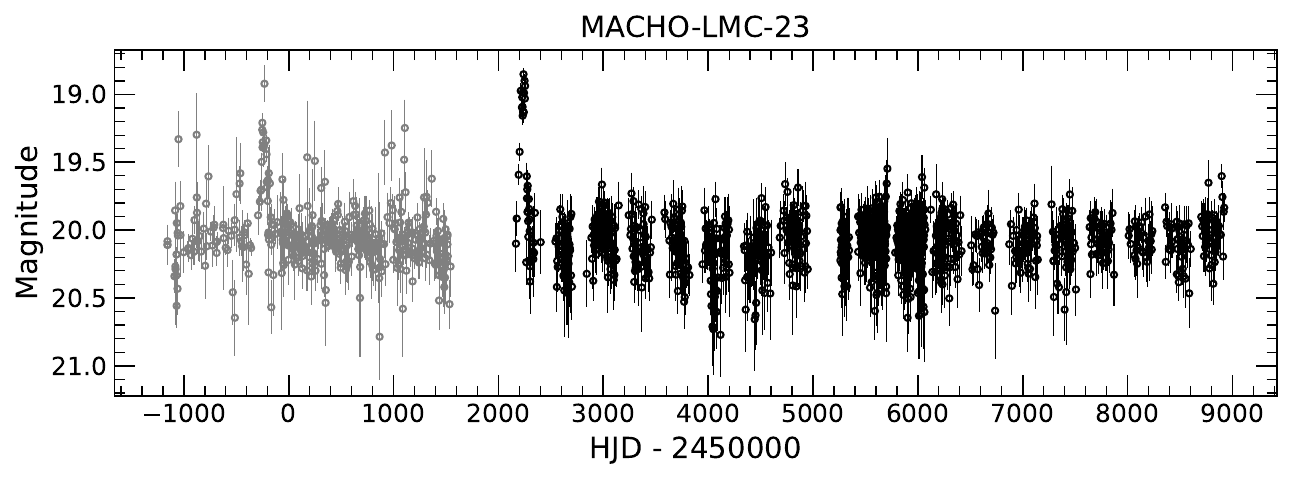}
\includegraphics[width=\textwidth]{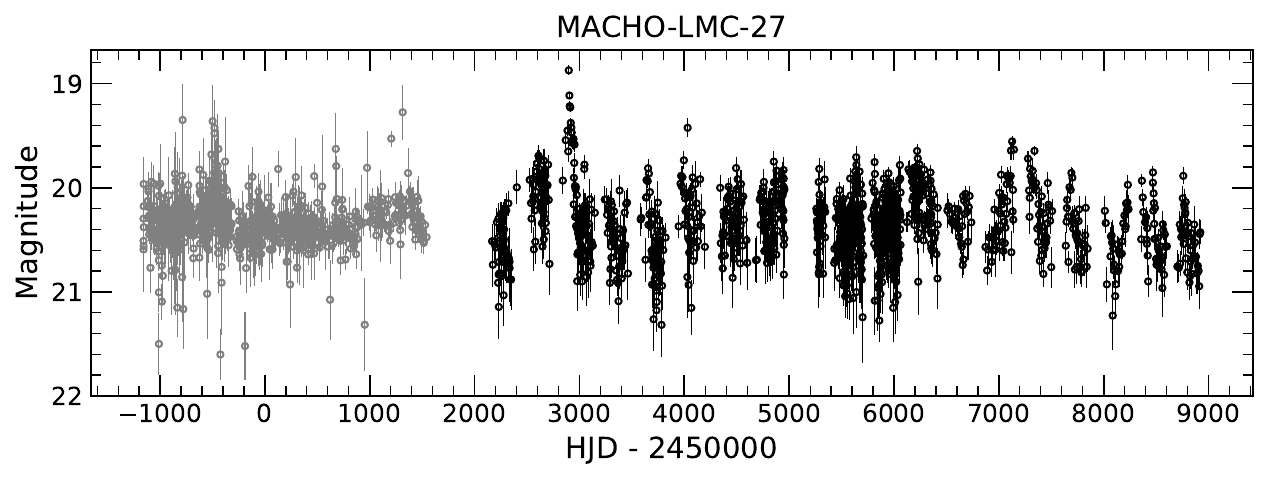}
\FigCap{Archival OGLE light curves  (black points) of MACHO-LMC-07, MACHO-LMC-23 and MACHO-LMC-27 reveal additional outbursts. The MACHO $R$-band data of the same stars are shown in gray. The MACHO magnitudes were shifted so their median matches that of the OGLE data.}
\label{fig:lc}
\end{figure}

As discussed in \citet{mroz2024a} (and previously noticed by several authors, e.g., \citealt{bennett2005b,tisserand2007,wyrzykowski2011}), two candidate events from the sample presented by \citet{alcock2000b} have exhibited additional outbursts in their archival light curves. Specifically, at least seven additional outbursts were recorded for MACHO-LMC-07 (in March 2005, January 2007, February 2009, November 2010, January 2012, November 2015, and September 2018) and one additional brightening was seen in the light curve of MACHO-LMC-23 (in November 2001). Additionally, MACHO-LMC-27 experienced one extra brightening in September 2003. The archival OGLE light curves for these three stars are shown in Fig.~\ref{fig:lc}.

Both MACHO-LMC-07 and MACHO-LMC-23 were recently identified by \citet{mroz2024c} as representatives of a new class of cataclysmic variables, dubbed as ``millinovae''. These objects exhibit repeating, triangle-shaped, symmetric optical outbursts, which seem to be associated with transient supersoft X-ray emission \citep{maccarone2019,mroz2024c}. Millinovae are likely close binary star systems consisting of a white dwarf accreting from an evolved donor star (a subgiant). Their orbital periods are of the order of a few days. In quiescence, their optical brightness is dominated by a large accretion disk, causing most millinovae to occupy a similar region in the color--magnitude diagram (CMD), approximately defined by $0.5 \lesssim (V-I)_0 \lesssim 1.0$, $1.5 \lesssim M_I \lesssim 2.5$ mag \citep{mroz2024c}.

\begin{figure}[htb]
\centering
\includegraphics[width=.7\textwidth]{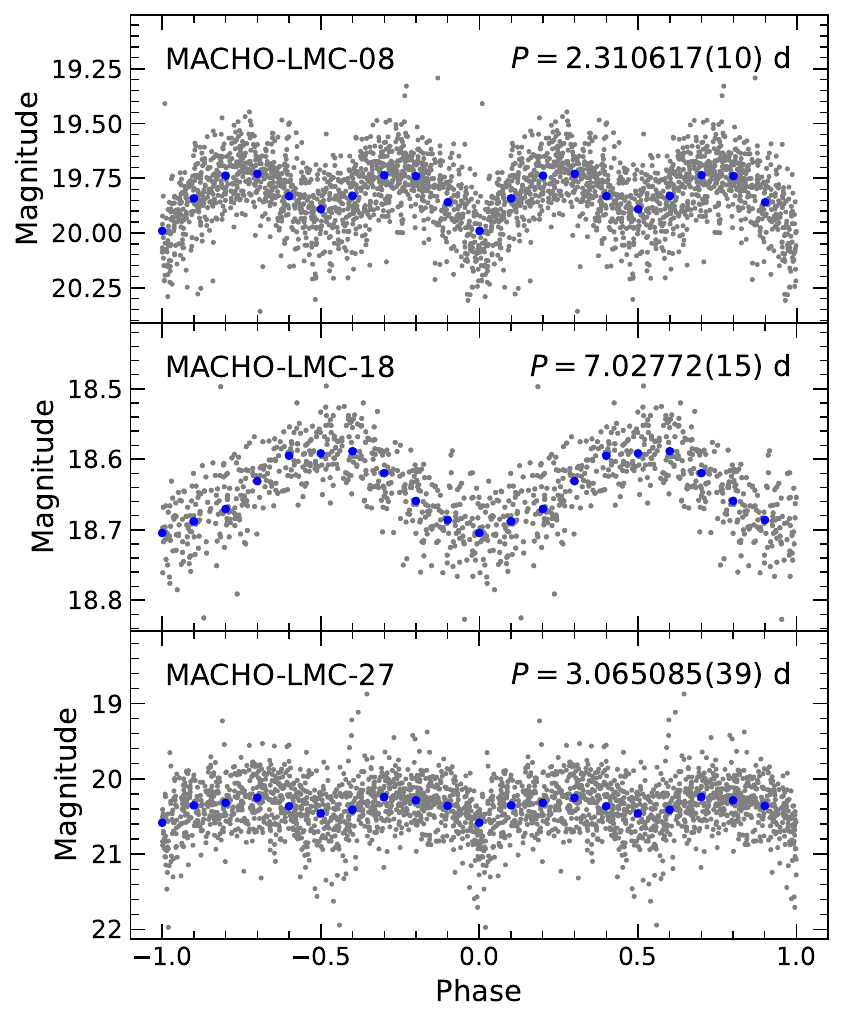}
\FigCap{Phase-folded OGLE light curves of three MACHO candidate events (MACHO-LMC-08, MACHO-LMC-18, and MACHO-LMC-27) that show periodic variability. The gray points mark the individual data points, and the blue points show the average light curve.}
\label{fig:lc2}
\end{figure}

In addition to identifying two outbursting systems, \citet{mroz2024a} found that three MACHO candidate events from \citet{alcock2000b} exhibit periodic photometric variability in their OGLE light curves. These are MACHO-LMC-08, MACHO-LMC-18, and MACHO-LMC-27, with variability periods of 2.310617(10), 7.02772(15), and 3.065085(39) days, respectively. (The orbital period of the latter system is twice as long as indicated in \citealt{mroz2024a}.) Their phase-folded light curves are presented in Fig.~\ref{fig:lc2}. The detected photometric variability suggests that the brightenings of MACHO-LMC-08, MACHO-LMC-18, and MACHO-LMC-27 likely had a stellar origin. While some microlensing events are known to have variable source stars \citep[e.g.,][]{wyrzykowski2006}, they are two orders of magnitude less frequent than microlensing events with constant sources. Therefore, finding three variable-source events in a sample of 13--17 events is extremely unlikely. Additionally, two of these events are located in a CMD region that is typically occupied by millinovae in quiescence. 

Removing these five variable objects (two showing repeating outbursts and three with periodic variability) from the sample of events used to derive the microlensing optical depth by \citet{alcock2000b} reduces it by almost 40\%, to $0.74 \times 10^{-7}$. 

\section{DIA Reductions of the MACHO Images}
\label{sec:dia}

Discussions and doubts regarding the status of the MACHO microlensing event candidates raised over the last decades, combined with the availability of images collected during the MACHO project, motivated us to take a closer look at the MACHO data with modern techniques. Thus, we have initiated a larger-scale program of revitalization of the MACHO photometry in the Magellanic Clouds and Galactic bulge fields. In the first step, we concentrated on extracting $R$-band light curves of microlensing candidates detected by MACHO toward the Magellanic Clouds. However, on a longer time scale, we plan to derive modern photometry for all objects detected in the MACHO images, mostly for archival purposes. The aim of this project is not to disavow the achievements of the MACHO project, which has made significant contributions to modern astrophysics, but to extend the life and legacy of the project for the next decades by providing revitalized MACHO data with the highest precision available today.

As described in the previous sections, the superiority of the DIA photometry technique, especially in the dense stellar fields, well proven by OGLE and other photometric surveys, encouraged us to attempt to apply this technique to the MACHO images. After several very promising tests, we have adjusted the OGLE photometry pipeline \citep{udalski2003} for the reduction of the MACHO images. While the details of this approach will be described elsewhere, here we present only the main highlights.

The MACHO mosaic camera for a given band consists of four $2048\times 2048$ CCDs, which are read by two amplifiers, resulting in eight $1024\times 2048$ pixel images of a given field. Because the position and reading directions of some CCDs in the mosaic were changed during the project, we decided to reduce the images in square subframes of $1024\times 1024$ pixels. Moreover, the MACHO telescope (The Great Melbourne Telescope at Mount Stromlo, Australia; unfortunately, it was burned and destroyed during the 2003 bush fires) provided reverse images on the East and West sides of the telescope pier. Thus, even when pointed at the same field, the same sky region could fall on different CCD detectors of the MACHO mosaic, depending on the telescope position.

To achieve a photometrically uniform data set over the whole duration of the MACHO project, we decided to use only one set of reference images for each $1024\times 1024$ pixel subframe. These reference images were constructed using the 5--10 best seeing and quality individual images collected on the East side of the pier. Then, all individual images for the DIA reductions were accordingly adjusted to the appropriate reference image depending on the CCD reading direction and the telescope position during the exposure (West or East side of the pier). In the cases of some detectors, rotation of the image by $\pm90^{\circ}$ was also necessary. Generally, the reference images were tied more to the specific areas of the sky rather than to the mosaic CCDs. We limited the set of MACHO images to those with seeing better than 7.2 MACHO pixels, \ie 4\zdot\arcs5 to maintain the best quality of photometry.

As expected, the comparison tests between the MACHO DIA photometry and the standard MACHO PSF photometry for several selected variable stars clearly showed a considerable improvement in quality. The light curves obtained using DIA photometry exhibited much less scatter and sharper features, enabling the detection of variability at much lower levels. Encouraged by the results of these tests, we extracted the DIA $R$-band photometry for all MACHO microlensing event candidates presented in samples A and B by \citet{alcock2000b}, with the exception of two candidates that were previously rejected because of repeating brightenings (MACHO-LMC-07 and MACHO-LMC-23). Additionally, we extracted the DIA light curve for the only MACHO SMC microlensing event candidate discovered by \citet{alcock1997d}.

Appendix~A presents the $R$-band light curves of the MACHO microlensing candidate events in two flavors, the original PSF MACHO light curve (retrieved from the MACHO archive) and the new DIA one. The full light curve of each candidate is shown in the main (left) panel, and a close-up of the detected brightening is shown in the right panel. The improvement in quality, especially for the fainter candidates, is remarkable. It should be noted that the variability of all candidates presented in Fig.~4 was easily detected in the new MACHO DIA photometry and the resulting light curves were of comparable quality to those of the OGLE ones in Fig.~4.

These more precise and detailed light curves allowed us to better qualitatively characterize the MACHO microlensing candidates. A summary of our findings is presented in Table~\ref{tab:class}.

\begin{table}
\begin{tabular}{lcl}
\hline
\hline
\multicolumn{1}{c}{Event} & \multicolumn{1}{c}{Sample} & \multicolumn{1}{c}{Comments} \\
\hline
MACHO-LMC-01  &   A  &  microlensing, anomaly at maximum,\\ 
&&possible binary-lens event \\
MACHO-LMC-04  &   A  &  microlensing \\
MACHO-LMC-05  &   A  &  microlensing \\
MACHO-LMC-06  &   A  &  weak candidate, asymmetric, nonphysical \\
&&blending in the microlensing fit \\
MACHO-LMC-07  &   A  &  many further brightenings \\
MACHO-LMC-08  &   A  &  periodic baseline, $P=2.310617$~d, asymmetric, \\
&&nonphysical blending in the microlensing fit \\
MACHO-LMC-09  &   B  &  binary-lens microlensing event \\
MACHO-LMC-13  &   A  &  weak candidate, asymmetric \\
MACHO-LMC-14  &   A  &  microlensing \\
MACHO-LMC-15  &   A  &  weak candidate, poorly covered maximum \\
MACHO-LMC-18  &   A  &  periodic baseline, $P=7.02772$~d, asymmetric \\
MACHO-LMC-20  &   B  &  possible microlensing, poorly covered event \\  
MACHO-LMC-21  &   A  &  microlensing, moderate coverage \\
MACHO-LMC-22  &   B  &  non-microlensing, fast rise, slow decline \\
MACHO-LMC-23  &   A  &  second brightening \\
MACHO-LMC-25  &   A  &  microlensing \\
MACHO-LMC-27  &   B  &  periodic baseline, $P=3.065085$~d, further \\
&&brightenings \\
\hline
MACHO-97-SMC-1 &   &   microlensing \\
\hline
\end{tabular}
\caption{Characterization of the MACHO events based on DIA photometry}
\label{tab:class}
\end{table}

Based on the new DIA light curves, we can identify seven objects (MACHO-LMC-(01, 04, 05, 09, 14, 21, 25)) as very likely microlensing events. While we do not attempt here to do a full analysis of the MACHO data and calculate the constraints on compact objects in dark matter based on data from this survey, we note that only these seven events would very likely pass any sensible microlensing cuts. Please note that our subjective selection is very consistent with that of \citet{belokurov2004}, having been further refined to exclude events with periodic variability in the baseline.

Two events from this sample (MACHO-LMC-01 and MACHO-LMC-09) are binary-lens events. The light curve of MACHO-LMC-01 shows an anomaly near the peak (as first suggested by \citealt{dominik1994,dominik1996}), while MACHO-LMC-09 is a caustic-crossing event \citep{bennett1996b}. This leaves us with only five single-lens single-source events, namely MACHO-LMC-(04, 05, 14, 21, 25), which are typically used for dark matter searches. It should also be recalled that the lens associated with MACHO-LMC-05 was directly detected several years after the event \citep{alcock2001b}, so that this event was caused by a nearby star located in the Galactic disk rather than a compact object in the Milky Way halo.

Objects MACHO-LMC-(07, 08, 18, 23, 27) are not genuine microlensing events, as indicated by additional brightenings and variability observed in their light curves. Their new DIA light curves (see Appendix~A) are asymmetric and do not match the expected microlensing profile. Furthermore, the light curve of  MACHO-LMC-22 is clearly asymmetric with a rapid rise followed by a slower decline. According to \citet{belokurov2004} and \citet{bennett2005}, this object is a Seyfert galaxy. The status of the four remaining candidates, MACHO-LMC-(06, 13, 15, 20), remains uncertain based on the analysis of the MACHO data alone. Their new DIA light curves are either asymmetric or lack sufficient coverage around the peak. Moreover, the best-fit microlensing models for these light curves converge to nonphysical values of the blending parameter, further suggesting that these are unlikely to be genuine microlensing events. On the other hand, \citet{bennett2005b} presented additional follow-up photometry around the maximum light for MACHO-LMC-13 and MACHO-LMC-15, which supports the microlensing interpretation.

\section{Selection of Microlensing Events}
\label{sec:selection}

\citet{hawkins2025} argued that OGLE is somehow biased against finding gravitational microlensing events in the photometric data. They brought up a number of possible explanations.

\subsection{Quality of Photometry}

The selection of events by \citet{mroz2024a} is based on the comparison of the light curve of a candidate event to the microlensing point-source point-lens model (also known as Paczy\'nski model).
\citet{hawkins2025} claimed that in the crowded star fields ``overlapping star images will distort the shape of the Paczy\'nski curve which is the primary criterion for identifying a microlensing event'' and ``changes in seeing from night to night will mean that source stars can remain single or merge at random with neighboring stars on a nightly basis, resulting in unpredictable changes to the Paczy\'nski profile''. 
These issues may have indeed posed challenges for the MACHO survey, which relied on PSF fitting photometry to extract light curves from science images. (However, it is important to emphasize that the MACHO team developed the best photometric pipeline they could to address these problems.) 

On the other hand, OGLE uses the DIA technique \citep{alard1998,wozniak2000} to extract light curves. Photometry is performed on difference (subtracted) images, which are created by subtracting a reference image from incoming frames. By doing this, the light from unrelated, non-variable stars is effectively removed, eliminating its influence on the measured light curves of microlensing events (see Fig.~\ref{fig:dia}). OGLE can achieve millimagnitude photometric precision, even in the most crowded stellar regions in the Galactic bulge and Magellanic Clouds \citep{udalski2015}. Fig.~\ref{fig:rms} demonstrates that the rms scatter in the OGLE light curves is a factor of a few lower compared to the MACHO light curves of the same stars. Therefore, the \citet{hawkins2025} claims that the ``features in a [OGLE] light curve are likely to be distorted by neighboring stars'' and that the OGLE light curves ``are inferior [to the MACHO data]'' are at best unsubstantiated. In fact, the high precision of OGLE photometry allows us to easily distinguish microlensing events from outbursting variable stars, supernovae, and other transient objects.

\subsection{Achromaticity}

According to \citet{hawkins2025}, ``observations in a single passband with no color information severely limit the likelihood of microlensing detection,'' and ``achromatic variation is an important requirement for identifying a microlensing event''. However, it has been common knowledge for decades that achromaticity is a second-order condition that is not necessary for classifying a brightening as a microlensing event. In the early, pioneering days of chasing for the first microlensing events, when photometry was derived using PSF fitting techniques, achromaticity played a much more important role in validating the reality of the event.

In addition, \citet{alcock2000b} analyzed relatively short light curves that spanned about 5.7\,yr. Thus, the condition of achromaticity of a candidate event was important to remove some outbursting sources from their sample. In contrast, because the duration of light curves analyzed by \citet{mroz2024a} is much longer (up to $\sim20$ years, plus an additional four years of OGLE-II photometry for some objects), repeating outbursts of some variable stars (such as millinovae, blue bumpers) can be easily recognized and removed from the sample of microlensing candidates.

Nevertheless, it is easy to envision a scenario where a source, such as a blue star, has an unresolved red companion. In this case, the color of the combined light measured with the PSF technique can change significantly when the source is magnified. Therefore, a genuine microlensing event may exhibit considerable chromaticity. Furthermore, the two-band observations conducted by the MACHO collaboration did not prevent \citet{alcock2000b} from including chromatic brightenings in their sample of LMC microlensing candidates (for example, MACHO-LMC-07 or MACHO-LMC-20). When discussing the discovery of a second outburst of MACHO-LMC-23, \citet{tisserand2007} concluded that ``original variation in the MACHO data is quite achromatic, indicating that achromaticity is not a fool-proof criterion for selecting microlensing events.''

The remedy for addressing blending and unresolved objects is the DIA technique, which was already discussed in the Section~\ref{sec:macho}. This method was pioneered by OGLE and has been successfully implemented in the OGLE pipeline already at the beginning of the 2000s. In this approach, the microlensing signal is measured on a difference image, where all neighboring and blended stars (blue, yellow, or red) are subtracted, as shown in Fig.~\ref{fig:dia}. As long as all the neighboring or blended stars within the radius of $\sim1.3$\,arcsec from the source are non-variable (which is the case in almost 100\% instances), the measured signal represents the genuine magnified light of the source, unaffected by any blended or unresolved stars.

Thus, right now, the main factor that counts for classifying a brightening as a microlensing event is its shape, specifically the fit to Paczy\'nski's curve in cases involving a single lens. Frequent observations in the second band are not necessary. Actually, some coverage of the event in the second band can be beneficial, for example, for the determination of the color of the source. However, it is possible to proceed without such observations. All major modern microlensing surveys (such as OGLE, MOA, KMTNet) collect the vast majority of their data in only one filter, and the generated light curves are used to search for microlensing events. Additional observations in the secondary filter are secured at much lower cadence, primarily to characterize the source star. 

\subsection{Cadence of Observations}

Another issue brought forward by \citet{hawkins2025} is the cadence of observations. They claim that the MACHO light curves are ``superior'' to those of OGLE in this respect. To illustrate their point, out of 16 microlensing events presented by \citet{mroz2024a}, they cherry-picked two events that happened to occur at the beginning or at the end of an observing season, when the cadence of observations is naturally lower.

\begin{figure}[htb]
\centering
\includegraphics[width=\textwidth]{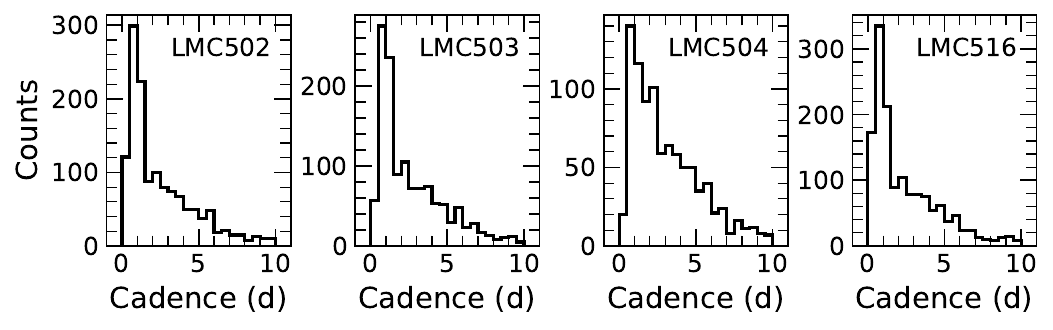}
\FigCap{Histograms of the observing cadence in four central OGLE LMC fields (LMC502, LMC503, LMC504, and LMC516).}
\label{fig:cadence}
\end{figure}

They also mention that the average OGLE cadence ``to be in the range 3--10 days''. However, this cadence varies depending on the field and the season. The central LMC fields, which contain the highest number of possible microlensing sources, were observed with a much higher cadence compared to the outer fields, which contain many fewer stars. Fig.~\ref{fig:cadence} shows histograms of the observing cadence in four central LMC fields (LMC502, LMC503, LMC504, and LMC516). The median cadence for LMC502, LMC503 and LMC516 is 2.0 days, while for LMC504, it is 3.0 days. This cadence is comparable to that of the MACHO survey (1.5--3.5 days according to \citealt{alcock2000b}), and it does not hinder the sensitivity to microlensing events.

\subsection{Bias against Blue Sources}

\citet{hawkins2025} claimed that OGLE is somehow biased against finding gravitational microlensing events with blue main-sequence stars. They argued that ``the source stars for the OGLE detections are almost entirely confined to the red giant branch, whereas the MACHO source stars are almost entirely confined to the top of the main sequence''. This claim is based on the analysis of the CMD for MACHO and OGLE events. First, \citet{hawkins2025} converted OGLE $I$-band magnitudes and $(V-I)$ colors into MACHO's $V$ and $(V-R)$ filters, although they did not provide any details regarding this transformation. Then, \citet{hawkins2025} read off the positions of MACHO events in the CMD from Fig.~7 of \citet{alcock2000b}. This figure shows the positions of blended (baseline) objects, which---as discussed above---may be influenced by blending effects. For example, MACHO-LMC-27 has $V-I=0.827$ measured from the OGLE reference images, but according to \citet{alcock2000b} it is much bluer ($V-R=0.07$) because it is blended with a nearby brighter blue star ($V=19.31$, $V-I=0.10$) in the MACHO template image.

We cross-matched all 17 microlensing event candidates identified by \citet{alcock2000b} with the OGLE databases to measure their mean magnitudes and colors. Thanks to the superior resolution of the OGLE reference images ($0.9\!-\!1.0''$), these measurements are less prone to systematics caused by crowding and blending compared to the MACHO template images ($1.5\!-\!2''$). Because the coordinates supplied in the archival MACHO data may be off by a few arcsec, we manually verified all the cross-identifications using MACHO finding charts and OGLE reference images. The mean magnitudes and colors are provided in Table~\ref{tab:macho}. We also listed the color excess $E(V-I)$ measured in the direction of each candidate event from the reddening maps of \citet{skowron2021}. We then calculated the de-reddened colors $(V-I)_0$ and magnitudes $I_0$ as follows:
\begin{align}
    \begin{split}
        I_0 &= I - 1.34 E(V-I), \\
        (V-I)_0 &= (V-I) - E(V-I),
    \end{split}
\end{align}
where we adopted the $I$-band extinction to reddening ratio of $A_I/E(V-I)=1.34$ from \citet{udalski2025}.

\begin{table}
\begin{tabular}{llrcccc}
\hline
\hline
\multicolumn{1}{c}{Event} & \multicolumn{1}{c}{Field} & \multicolumn{1}{c}{Number} & \multicolumn{1}{c}{$V$} & \multicolumn{1}{c}{$(V-I)$} & \multicolumn{1}{c}{$I$} & \multicolumn{1}{c}{$E(V-I)$} \\
\hline
MACHO-LMC-01  & LMC503.32 & 22975 & 19.722 & 1.065 & 18.657 & 0.099 \\
MACHO-LMC-04  & LMC502.13 & 3555 & 20.901 & 0.727 & 20.174 & 0.078 \\
MACHO-LMC-05  & LMC502.22 & 29743 & 20.717 & 1.608 & 19.109 & 0.086 \\
MACHO-LMC-06  & LMC502.17 & 60816 & 19.869 & 0.371 & 19.498 & 0.091 \\
MACHO-LMC-07  & LMC509.30 & 16531 & 21.233 & 0.647 & 20.586 & 0.083 \\
MACHO-LMC-08  & LMC503.01 & 27694 & 20.088 & 0.272 & 19.816 & 0.097 \\
MACHO-LMC-09  & LMC503.20 & 64924 & 20.785 & 0.664 & 20.122 & 0.066 \\
MACHO-LMC-13 & LMC503.26 & 51725 & 21.138 & 0.492 & 20.646 & 0.123 \\
MACHO-LMC-14 & LMC516.03 & 53635 & 19.497 & --0.006 & 19.503 & 0.097 \\
MACHO-LMC-15 & LMC509.21 & 48365 & 21.083 & 0.077 & 21.006 & 0.107 \\
MACHO-LMC-18 & LMC551.31 & 21415 & 19.321 & 0.929 & 18.392 & 0.161 \\
MACHO-LMC-20 & LMC530.27 & 29187 & 21.753 & 1.725 & 20.028 & 0.135 \\
MACHO-LMC-21 & LMC531.09 & 35833 & 19.464 & 0.189 & 19.275 & 0.153 \\
MACHO-LMC-22 & LMC501.31 & 27047 & 21.250 & 0.919 & 20.331 & 0.091 \\
MACHO-LMC-23 & LMC508.20 & 18608 & 20.930 & 0.796 & 20.134 & 0.124 \\
MACHO-LMC-25 & LMC511.07 & 2268 & 19.077 & 0.983 & 18.094 & 0.076 \\
MACHO-LMC-27 & LMC509.28 & 75025 & 21.239 & 0.827 & 20.412 & 0.134 \\
\hline
\end{tabular}
\caption{OGLE cross identification of MACHO candidate events.}
\label{tab:macho}
\end{table}

\begin{figure}[htb]
\centering
\includegraphics[width=.49\textwidth]{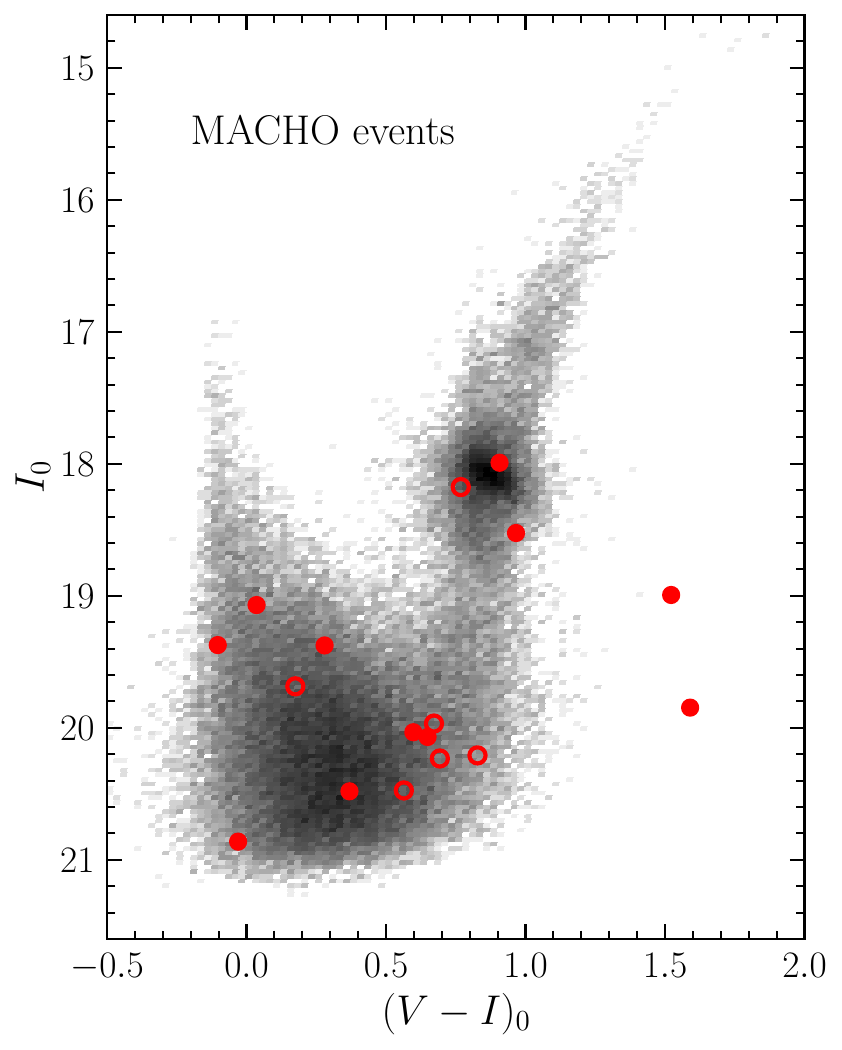}
\includegraphics[width=.49\textwidth]{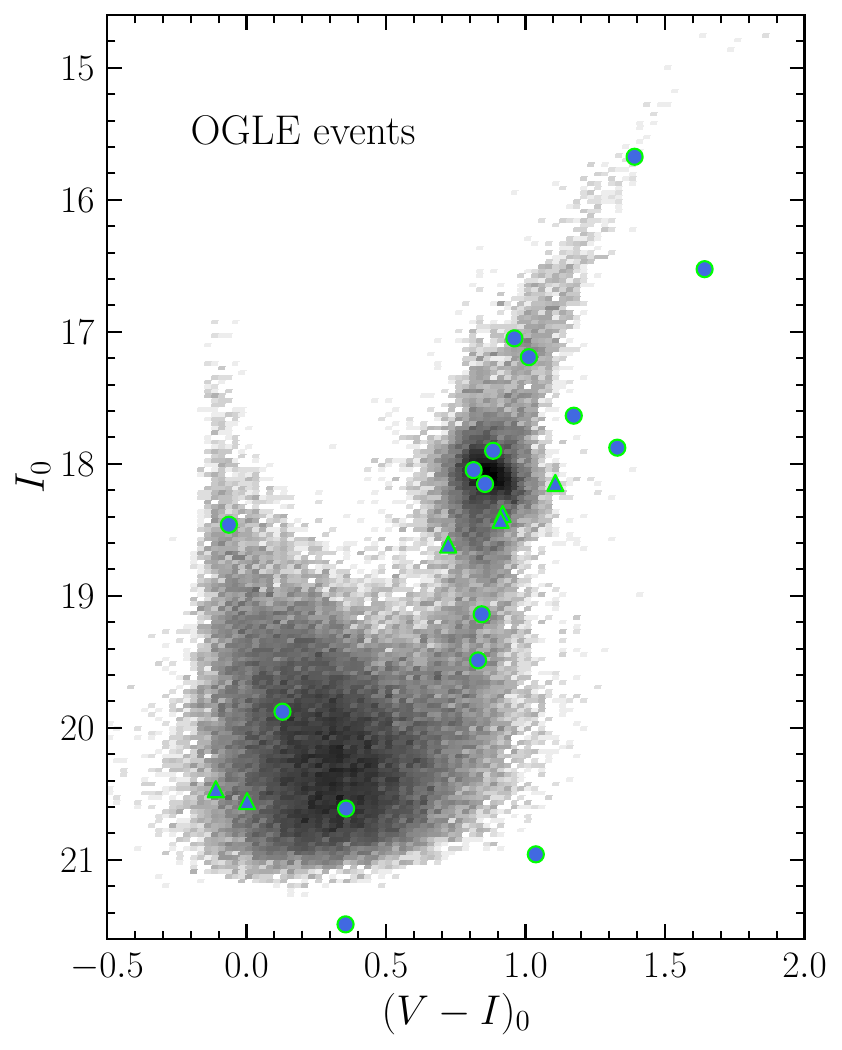}
\FigCap{Reddening-corrected color--magnitude diagrams for candidate microlensing events identified by \citet{alcock2000b} (left panel) and \citet{mroz2024a,mroz2025} (right panel). Open circles in the left panel mark five MACHO events with variable source stars and MACHO-LMC-22. Blue circles in the right panel show OGLE LMC events, blue triangles -- SMC events.}
\label{fig:cmd}
\end{figure}

The positions of the MACHO candidate events in the de-reddened CMD are shown in the left panel of Fig.~\ref{fig:cmd}. Five dubious events that exhibit additional photometric variability such as outbursts or periodic variations (Section~\ref{sec:macho}) and MACHO-LMC-22 (Section~\ref{sec:dia}) are marked with empty circles. Notably, five of six of these objects are located along the upper section of the main sequence, where MACHO is claimed by \citet{hawkins2025} to see an excess of events relative to OGLE.

The right panel of Fig.~\ref{fig:cmd} shows the locations of events found by \citet{mroz2024a} in the LMC (blue circles) and by \citet{mroz2025} in the SMC (blue triangles) in the de-reddened CMD. We subtracted 0.5\,mag from the magnitudes of the SMC events to take into account the difference in the distance modulus between the SMC and the LMC. As shown in this plot, events with blue main-sequence sources are not missing from our sample.

\subsection{Finite-Source Effects}

\citet{mroz2024a,mroz2024b} and \citet{wyrzykowski2009,wyrzykowski2011} selected microlensing events by fitting the standard point-source point-lens model to the light curves that exhibited any form of brightening from a flat baseline. They then evaluated several goodness-of-the-fit statistics. \citet{hawkins2025} argued that a ``substantial part of the LMC color--magnitude diagram is populated by (\dots) giants and supergiants, with radii ranging up to $3\times 10^{13}$\,cm, or around a third of the Einstein radius of the lens''. They implied that, in such a case, the light curve of a microlensing event would be distorted by finite-source effects and ``not recognized as a Paczy\'nski profile''.

Finite-source effects occur in high-magnification events when the lens crosses over the source's disk or when the angular Einstein radius of the lens is comparable to (or smaller than) the angular size of the source star. Consider a typical halo event with a $M=1\,M_{\odot}$ lens located at $D_l=20$\,kpc, and the source located at $D_s=50$\,kpc. The resulting angular Einstein radius is $\thetaE = \sqrt{\kappa M \pirel} \approx 0.5$\,mas, where $\kappa=8.144~\mathrm{mas}\,M_{\odot}^{-1}$, and $\pirel = \mathrm{au}/D_l - \mathrm{au}/D_s$ is the relative lens-source parallax. For self-lensing events, the angular Einstein radii are generally about an order of magnitude smaller, because the lenses are located close to the sources. Assuming $M=0.4\,M_{\odot}$, $D_l=48$\,kpc and $D_s=50$\,kpc, the angular Einstein radius is $\thetaE \approx 0.05$\,mas.

The angular size of a source star can be estimated using color--surface brightness relations.  For example, \citet{adams2018} found the following relationship for their ``all star'' sample:
\begin{equation}
\log(2\theta_*) = 0.542 + 0.391 (V-I)_0 - 0.2 I_0,
\label{eq:csb}
\end{equation}
where $\theta_*$ is the angular radius of the star (in mas), and $I_0$ and $(V-I)_0$ are de-reddened $I$-band brightness and $(V-I)$ color of that star, respectively. This relationship is valid for the color range $-0.05 \leq (V-I)_0 \leq 1.74$. 

\begin{figure}[htb]
\centering
\includegraphics[width=.49\textwidth]{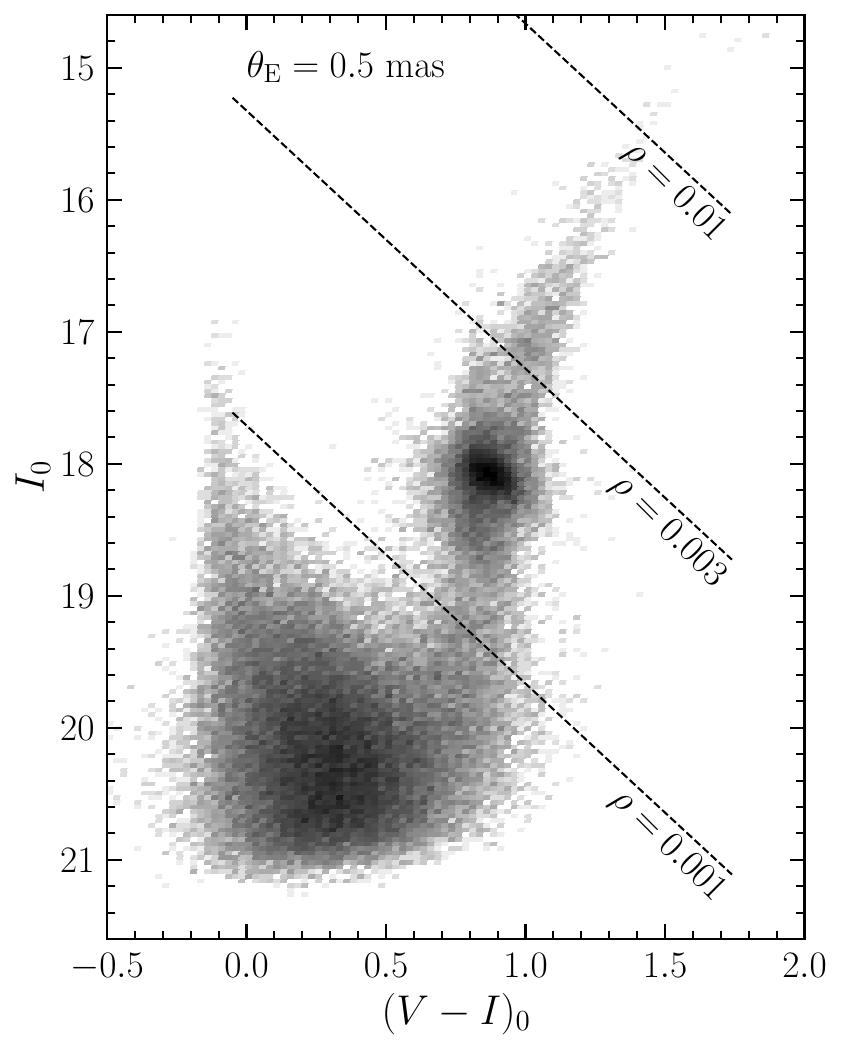}
\includegraphics[width=.49\textwidth]{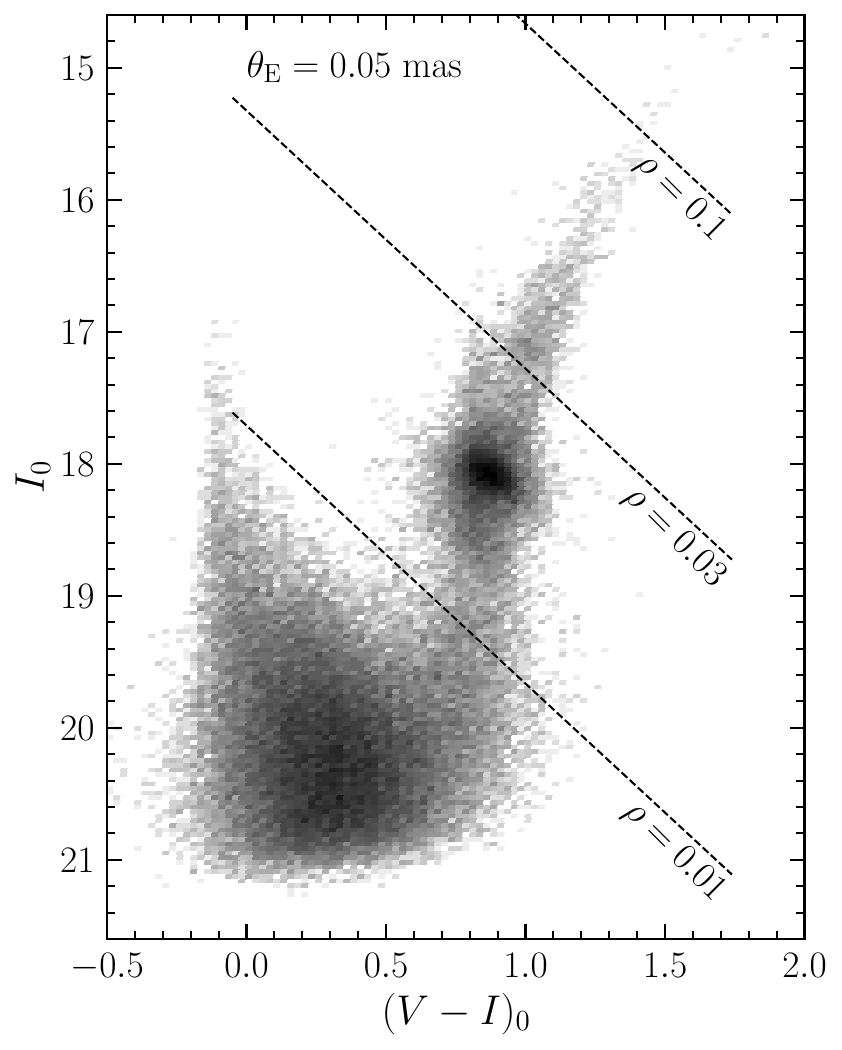}
\FigCap{Lines of constant $\rho$ calculated using the color--surface relation of \citet{adams2018} and assuming  $\thetaE=0.5$\,mas (microlensing by $1\,M_{\odot}$ PBHs; left panel) and $\thetaE=0.05$\,mas (self-lensing events; right panel). The background shows the de-reddened CMDs for the OGLE field LMC516.14. Because $\rho \ll 1$, finite-source effects can be neglected.}
\label{fig:finite}
\end{figure}

The finite-source effects are parameterized by the normalized angular radius of the source, defined as
\begin{equation}
\rho = \frac{\theta_*}{\thetaE}.
\label{eq:rho}
\end{equation}
Fig.~\ref{fig:finite} shows lines of constant $\rho$ calculated using Eq.~(\ref{eq:csb}) and (\ref{eq:rho}), under the assumption of two different Einstein radii: $\thetaE=0.5$\,mas (left panel) and $\thetaE=0.05$\,mas (right panel). The background shows the de-reddened CMDs for the OGLE field LMC516.14. For lensing by hypothetical $1\,M_{\odot}$ objects in the Milky Way halo (shown in the left panel of Fig. \ref{fig:finite}), finite-source effects are unimportant as $\rho<10^{-2}$ for virtually all possible source stars observed by OGLE. In cases of self-lensing events, finite-source effects may become important for the largest (brightest) sources with $I_0 \lesssim 15$ mag. However, such stars are exceedingly rare, constituting less than 0.3\% of all stars in the OGLE photometric databases. 
We also note that supergiants with radii of ``$2\times 10^{13}$\,cm'' \citep{hawkins2025} would have $I$-band magnitudes of about 13 and such stars are even two orders of magnitude more rare. Thus, neglecting finite-source effects does not affect the conclusions reached by \citet{mroz2024a,mroz2024b} and \citet{wyrzykowski2009,wyrzykowski2011}.

On the other hand, finite-source effects could become important only if we consider microlensing events by low-mass (planetary-mass) PBHs, with masses lower than $\approx 10^{-4}\,M_{\odot}$. In fact, OGLE carried out a special high-cadence survey of the Magellanic Clouds with the goal of searching for short-timescale microlensing events that may be caused by such low-mass PBHs if they existed \citep{mroz2024d}. However, we found no compelling short-timescale events \citep{mroz2024d}. The finite-source effects were taken into account both when searching for microlensing events in the data from the high-cadence survey and when calculating the event detection efficiency.

\subsection{Supernovae}

\begin{figure}[htb]
\centering
\includegraphics[width=\textwidth]{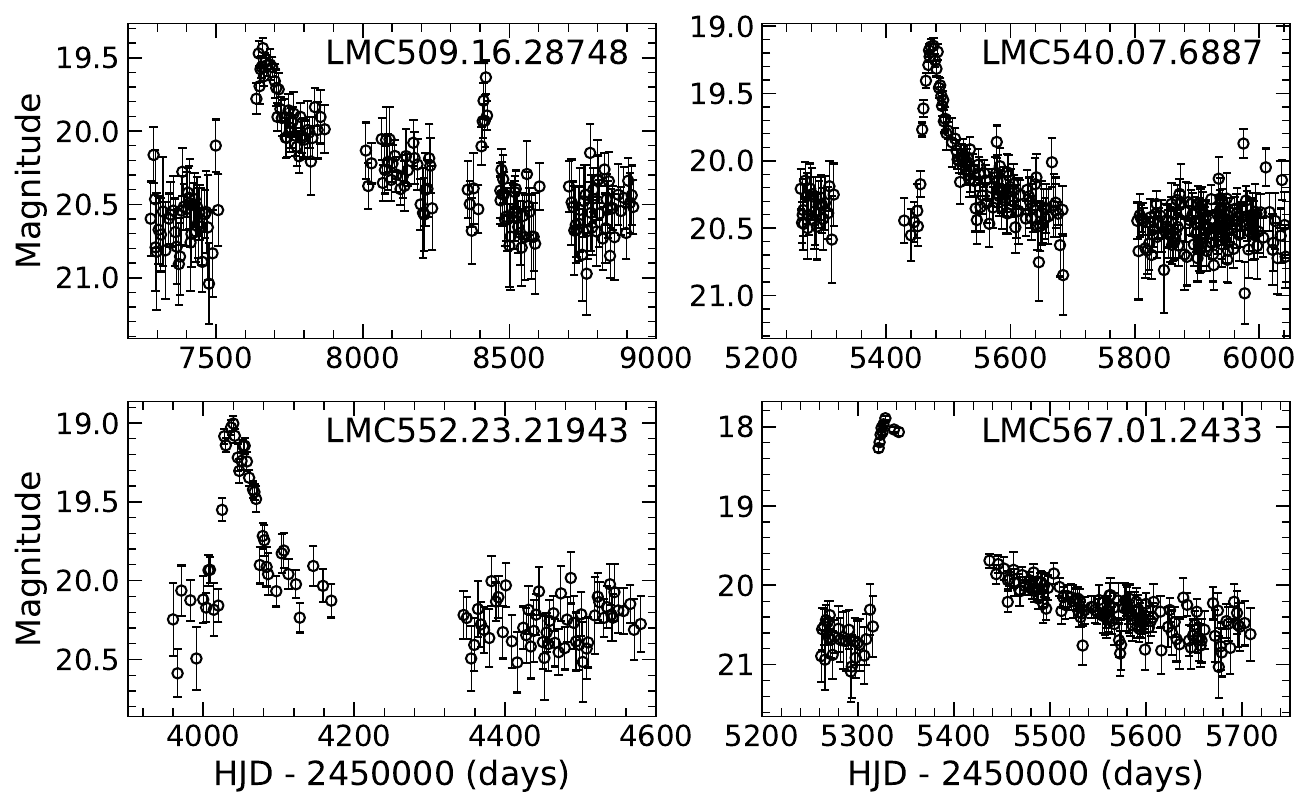}
\FigCap{Example light curves of supernovae and other extragalactic transients discovered in the OGLE data, which did not pass the selection criteria for microlensing events.}
\label{fig:sne}
\end{figure}

\citet{hawkins2025} argued that ``the MACHO group use color information to help identify supernovae light curves, a source of contaminant which does not appear to be mentioned by \citet{mroz2024a}''. Supernovae were not discussed in that study because they did not contaminate our sample of events. First, we analyzed only the light curves of stellar sources that were detected in the reference image. Second, supernova light curves were excluded by the selection criteria devised by \citet{mroz2024a}. However, we went through a list of objects that were selected in the first stage of our algorithm, and found many supernovae and other extragalactic transients that did not make it to the final sample of microlensing events. Fig.~\ref{fig:sne} shows some example light curves. A dedicated survey for supernovae and other transients in data collected by the OGLE project is presented by \citet{wyrzykowski2014}.

\subsection{Variability}

\citet{hawkins2025} noticed that ``part of the OGLE event selection procedure is to reject any event where the source shows signs of intrinsic variability'' and claimed that ``this will result in a high probability that microlensing events in their light curves will be ignored''.

As part of the event selection process, we calculated the following quantity:
\begin{equation}
    \chi^2_{\rm out}=\sum_{i=1}^N \frac{(F_i - F_{\rm base})^2}{\sigma_i^2},
\end{equation}
where $F_{\rm base}$ is the mean flux in the baseline (unmagnified part) of the event and ($F_i$,$\sigma_i$) are the measured flux and its uncertainty during the $i$th epoch. The summation is performed over all data points located outside the time window containing the candidate event (this ensures that the magnified part of the light curve does not contribute to $\chi^2_{\rm out}$). We required that $\chi^2_{\rm out}/(N-1)$ to be less than 2. This selection cut removes only a few percent of the most variable sources.

\begin{figure}[htb]
\centering
\includegraphics[width=.9\textwidth]{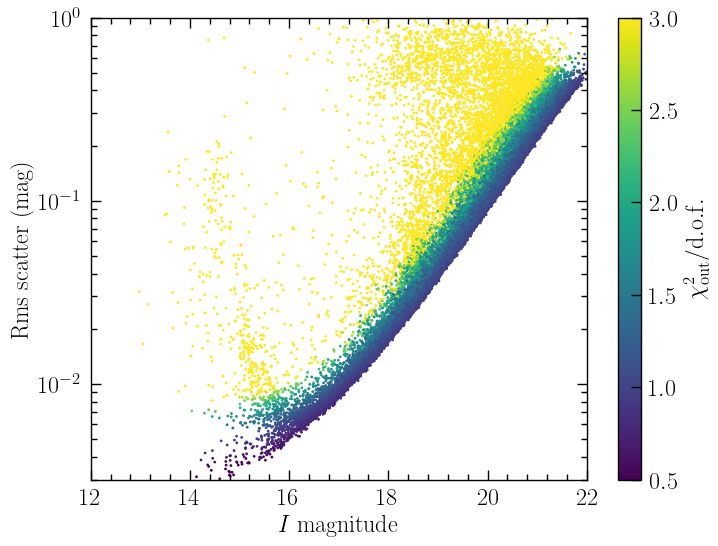}
\FigCap{Rms scatter as a function of the mean magnitude for stars in the field LMC516.14. The data points are color-coded by a value of the statistic $\chi^2_{\rm out}/(N-1)$. When searching for microlensing events we discarded objects with $\chi^2_{\rm out}/(N-1)>2$, which comprise only 6\% of all stars in the field.}
\label{fig:baseline}
\end{figure}

To illustrate this, we calculated $\chi^2_{\rm out}/(N-1)$ for all stars located in a randomly selected field LMC516.14. We also calculated the rms scatter in their light curves. Fig.~\ref{fig:baseline} shows the rms scatter as a function of the mean magnitude of the stars. Each data point is color-coded according to its respective value of $\chi^2_{\rm out}/(N-1)$. As can be seen from this figure, the $\chi^2_{\rm out}/(N-1) \leq 2$ cut eliminates only a small number of objects ($\approx 6\%$) that exhibit the largest variability, while retaining 94\% of the stars in the sample. Therefore, the claims made by \citet{hawkins2025} are simply incorrect.

\begin{figure}[htb]
\centering
\includegraphics[width=.9\textwidth]{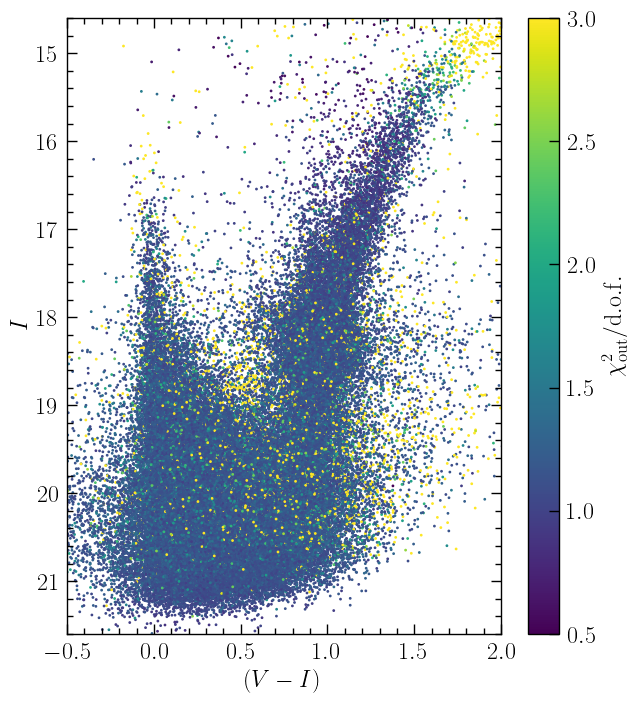}
\FigCap{Color--magnitude diagram of the field LMC516.14. The data points are color-coded by a value of the statistics $\chi^2_{\rm out}/(N-1)$.}
\label{fig:variability}
\end{figure}

As an additional check, we plotted the CMD of the field LMC516.14 in Fig.~\ref{fig:variability}. Each data point is color-coded according to the value of the statistic $\chi^2_{\rm out}/(N-1)$. We found that objects rejected because of their variability are distributed approximately uniformly across the CMD. There are two regions with an increased density of variable stars. The first region is located in the upper part of the red giant branch ($I \lesssim 15.5$ mag, $(V-I) \gtrsim 1.5$), where red giant pulsations dominate the photometric variability of stars. As explained in the previous section, such stars are very rare (less than 0.3\% of all stars in the databases). Second, there appears to be an overdensity of variable stars between the main sequence and the red giant branch around $0.4 \lesssim (V-I) \lesssim 0.6$, $18.6 \lesssim I \lesssim 19.0$ mag, which is coincident with the pulsation instability strip, occupied by Cepheids and RR Lyrae stars.

Finally, \citet{hawkins2025} claimed that ``most of the OGLE sample of microlensing candidates lie in the red giant part of the color--magnitude diagram, and are likely to be variable''. While it is true that many microlensing events presented in the study of \citet{mroz2024a} have subgiant and giant sources, the inspection of the residuals from the best-fit microlensing models (see Fig.~4 of \citealt{mroz2024a}) does not reveal any significant variability, either periodic or otherwise. The rms scatter in the light curves is consistent with that of other constant stars of similar magnitude in the field. Of course, all stars can be considered variable when studied at sufficiently high precision, such as with space telescopes, but there is no evidence of their variability in the OGLE data.

\subsection{Selection Criteria}

The final comment by \citet{hawkins2025} is related to the selection criteria. The MACHO collaboration \citep{alcock2000b} presented two sets of selection cuts: one aimed at selecting only high-quality light curves, while the other was less stringent. In their work, \citet{mroz2024a} used only one set of selection criteria, which led \citet{hawkins2025} to raise concerns that this approach may have missed many microlensing events. 

To address this issue, \citet{mroz2024a} and \citet{wyrzykowski2009,wyrzykowski2011} not only applied the automated selection criteria but also conducted a visual inspection of all the light curves that passed the initial selection cut. This additional step was taken to ensure that no large number of genuine microlensing events were missed by the algorithm. A trained human eye can easily recognize anomalous events, such as those exhibiting finite-source effects or binary-lens events, which the algorithm may have left out. \citet{mroz2024a} reported finding of only three additional events using by-eye searches, confirming the high completeness of the sample selected according to the objective selection criteria. Furthermore, the selection criteria were developed using simulated light curves of microlensing events, with the goal of maximizing the event detection efficiency while keeping the contamination from non-microlensing light curves as minimal as possible.

\section{Detection Efficiency}
\label{sec:efficiency}

Knowledge of the event detection efficiency, which is a function of the event timescale, is an integral part of the calculation of the microlensing optical depth and determination of the limits on the contribution of compact objects to dark matter. The methodology of calculating the event detection efficiency in the work of \citet{mroz2024a} necessarily borrows from our extensive experience in such computations based on the analysis of the OGLE data in the Galactic bulge \citep{mroz2017,mroz2019}.

Broadly speaking, there are two possible approaches for calculating the event detection efficiency. First, one can inject synthetic microlensing events into the science images of the survey, then process all the data using the same photometric and event detection pipeline as used for the real data. An example of such calculations is presented in \citet{mroz2017}. Another possibility is to inject microlensing signals into the existing light curves of stars in a given field. \citet{mroz2019} presented a detailed description of such an algorithm, which is designed to preserve variability and noise that are present in the original light curves. Furthermore, \citet{mroz2019} demonstrated that both approaches (image-based and light-curve-based simulations) yield nearly identical results in the crowded Galactic bulge fields (see Fig.~10 of \citealt{mroz2019}). These fields are a factor of a few more crowded than the LMC and SMC fields analyzed in the works of \citet{mroz2024a,mroz2025}.

\citet{hawkins2025} correctly argue that the robustness of the detection efficiency simulations can be verified ``where there is an overlap between two surveys when the same field is observed by both surveys at the same time''. While the overlap between OGLE and MACHO observations of the LMC is very small, our approach for determining the event detection efficiency can be independently validated by comparing the microlensing optical depth and event rate measurements in the Galactic bulge. Such measurements were presented by \citet{mroz2019}, based on eight years of OGLE observations of the Galactic bulge, and by \citet{nunota2025}, based on nine years of observations by the Microlensing Observations in Astrophysics (MOA) survey. Fig.~3 of \citet{nunota2025} demonstrates that these quantities, independently measured by OGLE and MOA teams, match very well. This test provides independent validation of our light-curve-based calculations of the event detection efficiency in the Galactic bulge fields (which, we emphasize, may be an order of magnitude more crowded than the Magellanic Clouds fields). Because the work of \citet{mroz2024a} used essentially the same methodology for calculating the detection efficiency as that of \citet{mroz2019}, there is no evidence that our detection efficiencies are incorrect.

\subsection{Distributions of $u_0$ and $t_0$}

\begin{figure}[htb]
\centering
\includegraphics[width=.49\textwidth]{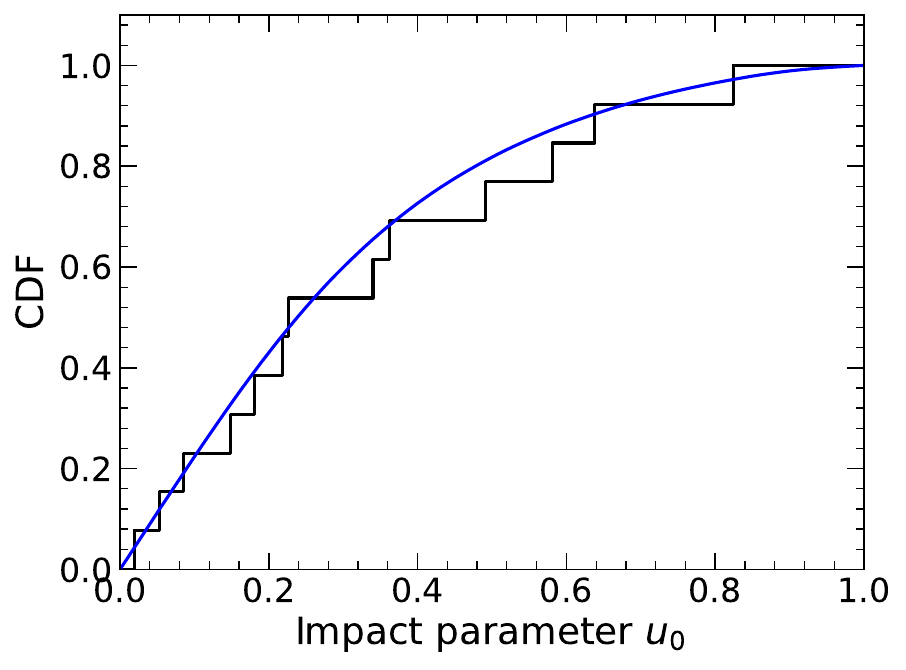}
\includegraphics[width=.49\textwidth]{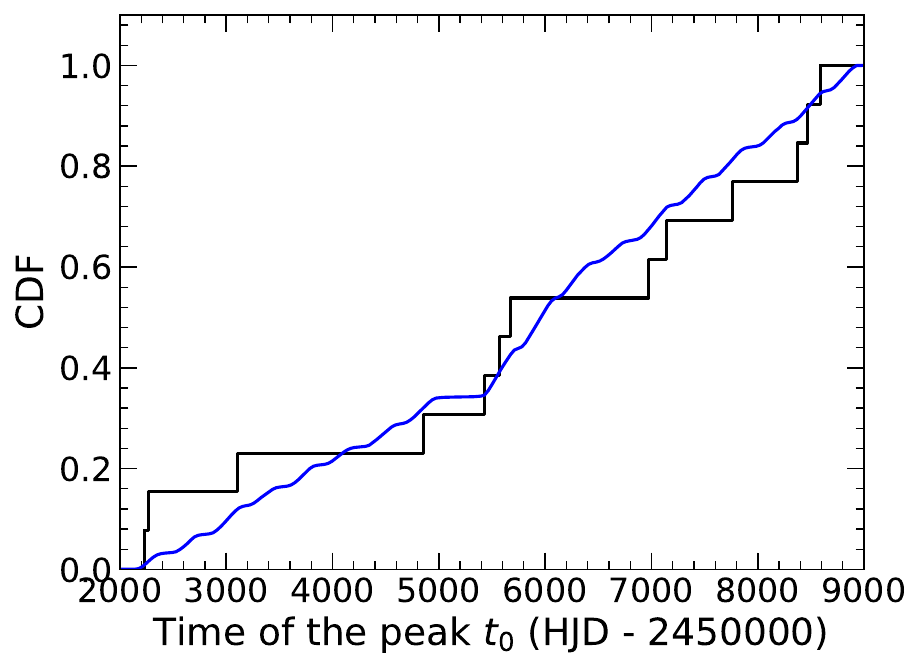}
\FigCap{Cumulative distributions of impact parameters (left panel) and moments of maxima (right panel) of events included in the statistical sample of \citet{mroz2024a,mroz2024b}. The blue lines show the expected distributions calculated using simulated light curves. Observed and expected distributions match very well, with the $p$-values of the Kolmogorov--Smirnov test of 0.983 and 0.938 for $u_0$ and $t_0$ distributions, respectively.}
\label{fig:cdf}
\end{figure}

Another method that enables us to validate the results of detection efficiency simulations is to compare the expected distributions of some parameters with their observed values. The left panel of Fig.~\ref{fig:cdf} shows the cumulative distribution of events' impact parameters $u_0$, while the right panel presents the cumulative distribution of the moments of maxima $t_0$. For the observed distributions of these parameters, we adopted the best-fit values provided in Tab.~6 of \citet{mroz2024a}. For events with degenerate parallax solutions, we adopted the parameters corresponding to the $u_0>0$ solution. We only considered 13 events that met the objective selection criteria and were included in the statistical sample of events of \citet{mroz2024a,mroz2024b}.

The distributions of $u_0$ and $t_0$ should be, in theory, uniform. However, we are more likely to select events with certain values of $u_0$ and $t_0$. For example, the amplitude of an event depends on the value of the impact parameter (a smaller impact parameter corresponds to a higher amplitude). As a result, we are more sensitive to events with lower values of $u_0$. Similarly, we are more likely to select events that peak mid-season. Events that occur during seasonal gaps, or at the very beginning or end of the season, are less likely to be found and included in our statistical sample.

We used the results of simulations presented by \citet{mroz2024a} to calculate the expected cumulative distributions of the parameters $u_0$ and $t_0$ in each observed OGLE field. In these simulations, the peak time $t_0$ was drawn from a uniform distribution within the following ranges: $2452000 \leq t_0 \leq 2459000$ for fields observed during the \mbox{OGLE-III} and \mbox{OGLE-IV} phases, and $2455000 \leq t_0 \leq 2459000$ for fields observed during \mbox{OGLE-IV} only. The impact parameter was drawn from a uniform distribution from the range $0 \leq u_0 \leq 1$. Because our original goal was to calculate the event detection efficiency as a function of the event timescale, timescales were drawn from a log-uniform distribution from the range 1 to $10^4$ days.

However, the timescales of actually observed events (i.e., those expected from stars in the Magellanic Clouds and the Milky Way disk) follow a different distribution \citep{mroz2024b}. To address this, we assigned each simulated event a weight calculated as follows:
\begin{equation}
    w = \exp\left(-\frac{(\log\tE - 1.6)^2}{2 \cdot 0.4^2}\right).
\end{equation}
This weight was chosen to match the timescales of expected events. However, we found that our conclusions do not strongly depend on the adopted weighting scheme. Even when we applied no weights at all, the distributions of $u_0$ and $t_0$ remained similar to those presented in Fig.~\ref{fig:cdf}. 

Subsequently, we calculated the event detection efficiency as a function of $u_0$ and $t_0$ in each field, which is defined as the fraction of simulated events that satisfy the criteria presented in Tab.~4 of \citet{mroz2024a}. For $u_0$, we divided the range from 0 to 1 into 100 identical bins. For $t_0$, we used 200 identical bins spanning from 2452000 to 2459000. We then calculated the average detection efficiency across all OGLE fields, assigning each field a weight equal to the product of the number of source stars located within it and the duration of its observations. 

The resulting cumulative distribution functions for $u_0$ and $t_0$ are presented by blue lines in Fig.~\ref{fig:cdf}. To assess how well the data match the expected distributions derived from our simulations, we performed the Kolmogorov--Smirnov test. We obtained $p$-values of 0.983 and 0.938 for the distributions of impact parameters and peak times, respectively. Therefore, the observed distributions of $u_0$ and $t_0$ are consistent with those expected from our simulations.

According to \citet{hawkins2025}, ``it is perhaps strange that the earlier phase of the OGLE program, although lasting for 13 years only detected 4 microlensing events, whereas the final phase lasting 10 years detected 13 events in roughly the same area of the sky''. This statement is incorrect, unfortunately. First of all, by ``earlier phase of the OGLE program'', \citet{hawkins2025} probably mean the \mbox{OGLE-II} (1996--2000) and \mbox{OGLE-III} phases (2001--2009) and the works of \citet{wyrzykowski2009} and \citet{wyrzykowski2011}. As explained in Section~\ref{sec:macho}, in total only three genuine events were identified during these phases. Both \mbox{OGLE-II} and \mbox{OGLE-III} covered a significantly smaller area than the latter phase of the survey (\mbox{OGLE-IV}).

On the other hand, \citet{mroz2024a,mroz2024b} searched for microlensing events in the combined \mbox{OGLE-III} and \mbox{OGLE-IV} light curves, which span nearly 20 years from 2001 to 2020. They reported the discovery of 13 events, four of which occurred during the \mbox{OGLE-III} phase (2001--2009) and nine during the \mbox{OGLE-IV} phase (2010--2020). As shown in the right panel of Fig.~\ref{fig:cdf}, based on our simulations, we expect that about 34\% of all selected events (that is, around 4.4) should have happened before 2010, in excellent agreement with observations.
The higher rate of event detections in \mbox{OGLE-IV} than in \mbox{OGLE-III} can be explained by two factors: the larger number of observed stars and the higher detection efficiency.

\subsection{OGLE and MACHO Overlap}

Finally, when discussing detection efficiencies, \citet{hawkins2025} complained that ``there is no published record of any detections by the OGLE group of MACHO microlensing candidates''. This is not surprising given that two surveys did not overlap much. \citet{alcock2000b} reported the results of the analysis of MACHO data collected between 1992 July 22 and 1992 August 22, and from 1992 September 18 to 1998 March 17. On the other hand, the \mbox{OGLE-II} data analyzed by \citet{wyrzykowski2009} cover the period from 1996 December 28 (and observations of some fields started later than that) to 2000 November 25. Only two MACHO event candidates, MACHO-LMC-14 and MACHO-LMC-15, were located in the \mbox{OGLE-II} fields and their brightness was declining from the peak during the period of \mbox{OGLE-II} observations. Because these events had already peaked before \mbox{OGLE-II} began its observations, they could not be confidently selected as microlensing candidates by \citet{wyrzykowski2009}. The OGLE light curves of other microlensing candidates identified by \citet{alcock2000b} are presented in Section~\ref{sec:macho}.

\section{Self-lensing and Milky Way Dark Matter Halo Models}
\label{sec:models}

\subsection{Self-lensing and Lensing by Stars in the Milky Way Disk}

The microlensing optical depth toward a given source star at a distance $D_s$ is given by the formula:
\begin{equation}
    \tau(D_s) = \frac{4\pi G}{c^2}\int_0^{D_s} \rho(D_l) \frac{D_l(D_s-D_l)}{D_s}dD_l,
    \label{eq:tau}
\end{equation}
where $D_l$ is the distance to the lens and $\rho(D_l)$ is the density of lenses along the line of sight \citep[e.g.,][]{paczynski1996}. However, the microlensing optical depth measured by an actual survey is the average over the population of the source stars observed by that survey:
\begin{equation}
    \tau = \frac{\int \tau(\alpha,\delta,D_s) dn(\alpha,\delta,D_s)}{\int dn(\alpha,\delta,D_s)},
\end{equation}
where $(\alpha,\delta)$ are the equatorial coordinates. Consequently, even if we assume the same model for the distribution of lenses, the microlensing optical depth averaged over \mbox{OGLE-II}, \mbox{OGLE-III}, and \mbox{OGLE-IV} footprints will be different. This discrepancy is illustrated in Fig.~1 of \citet{calchi_novati_2009}, who shows that the microlensing optical depth averaged over the central LMC regions is higher than in the outer regions. Thus, comparing the optical depths measured by \mbox{OGLE-III} \citep{wyrzykowski2011} and \mbox{OGLE-IV} \citep{mroz2024a} with that calculated in the MACHO fields \citep{alcock2000b,mancini2004}, as \citet{hawkins2025} did, is misleading.

A more robust approach is to calculate the expected number of microlensing events in a specific field, taking into account the distribution of positions and velocities of lenses and sources, event detection efficiencies, and the total number of source stars detected. This approach is used, for example, by \citet{calchi_novati_2009} and \citet{calchi_novati_2011}, who---using the same LMC model---predicted a total of 0.64 and 1.37 microlensing events in the \mbox{OGLE-II} and \mbox{OGLE-III} bright star samples, respectively. These numbers are in good agreement with the observed one event in the \mbox{OGLE-II} data\footnote{\citet{wyrzykowski2009} originally reported two candidate events in the \mbox{OGLE-II} data. However, one of them, OGLE-LMC-02, later turned out not to be a microlensing event.} \citep{wyrzykowski2009} and two events in the \mbox{OGLE-III} data \citep{wyrzykowski2011}.

The same approach was also adopted by \citet{mroz2024b}. According to their model, a total of 12.7--20.4 events caused by known stellar populations should be detected in the combined \mbox{OGLE-III}/\mbox{OGLE-IV} data set, depending on the adopted Milky Way disk model. These numbers should be compared to the total of 13 microlensing events in the statistical sample of \citet{mroz2024a,mroz2024b}. \citet{hawkins2025} quote the self-lensing optical depth of $\tau=(0.4 \pm 0.1) \times 10^{-7}$ toward the LMC based on models published by \citet{alcock2000b}, \citet{mancini2004}, and \citet{calchi_novati_2009}. These values were calculated for the inner regions of the LMC to match the footprints of the MACHO and \mbox{OGLE-II} surveys. Thus, for the reasons described above, they are not applicable to the combined \mbox{OGLE-III}/\mbox{OGLE-IV} data, which cover a significantly larger area than the previous surveys. Furthermore, \citet{hawkins2025} claim that ``the expected number of around 40 self-lensing events'' should have be detected by \citet{mroz2024a}. However, they did not provide any calculation or justification of this number.

\citet{hawkins2025} argued that the light of stellar lenses located in the Milky Way disk should be detectable in the light curves (and perhaps spectra) of microlensing events. While we do not think that detecting light from a ``K5 main-sequence star at 25 kpc'' in the microlensing event light curve is possible, half of the events reported by \citet{mroz2024a} show evidence for blended light, which most likely originates from the lens itself.

For example, the $I$-band brightness of the blend in OGLE-LMC-04 is $I=18.2^{+1.1}_{-0.5}$ mag, OGLE-LMC-07: $I=18.08^{+0.47}_{-0.07}$  mag, OGLE-LMC-09: $I=19.30^{+0.83}_{-0.41}$ mag, OGLE-LMC-10: $I=19.33^{+0.56}_{-0.33}$ mag, OGLE-LMC-13: $I=17.9^{+1.1}_{-0.5}$ mag, OGLE-LMC-14: $I=19.67^{+0.88}_{-0.26}$ mag, OGLE-LMC-17: $I=18.2^{+1.3}_{-0.7}$ mag, OGLE-LMC-18: $I=20.87^{+0.07}_{-0.04}$ mag. 

We calculated the surface density of stars detected on the OGLE reference images that are brighter than the blend. This allowed us to estimate that the probability of finding a random star brighter than the blend within the seeing disk of the source is $4.0\%$, $6.8\%$, $5.0\%$, $5.8\%$, $4.3\%$, $11.3\%$, $0.7\%$ for OGLE-LMC-(04, 07, 09, 10, 13, 14, 17), respectively. These numbers indicate that the blended light is unlikely to originate from unrelated stars, but most likely comes from a luminous lens. The number of events showing significant blended light is consistent with the expected number of events from stars int the Milky Way \citep{mroz2024b}.

\subsection{Milky Way Halo}

Finally, \citet{hawkins2025} argue that the Milky Way halo model used by \citet{mroz2024b} to infer limits on the PBH content in dark matter is incorrect. \citet{mroz2024b} adopted the contracted halo model of \citet{cautun2020}, which was inferred by fitting physically motivated models to the \textit{Gaia} DR2 Galactic rotation curve \citep{eilers2019} and other observational data. This model estimates the total mass of the Milky Way dark matter halo to be $0.41 \times 10^{12}\,M_{\odot}$ within 50 kpc of the Milky Way center, predicting a circular velocity of around 190 km\,s$^{-1}$ at that distance.

Some more recent works \citep[e.g.][]{wang2023,jiao2023,ou2024}, based on the newer \textit{Gaia} DR3 data, seem to indicate that the rotation curve of the Milky Way is declining at Galactocentric distances greater than 15--20\,kpc. That implies that the Milky Way mass is smaller than previously thought (with the total dark matter halo mass estimated to be about $0.21 \times 10^{12}\,M_{\odot}$ in the \citet{jiao2023} model and about $0.18 \times 10^{12}\,M_{\odot}$ in the \citet{ou2024} model). These studies derived the rotation curve using the Jeans equation. However, some follow-up studies argued that disequilibrium and deviations from axisymmetry could lead to large systematic errors in the rotation curves inferred using Jeans equation \citep[e.g.,][]{koop2024,klacka2025}.

Nevertheless, as discussed by \citet{mroz2024b}, the choice of the Milky Way model does not significantly affect their conclusions regarding the limits on the abundance of PBHs in dark matter (see their Extended Data Figure~2). This is because the inferred limits on the PBHs abundance are inversely proportional to the expected number of microlensing events by hypothetical compact objects in the dark matter halo. This number is, in turn, directly proportional to the microlensing optical depth of the Milky Way halo in a given direction. 

For the B2 halo model of \citet{jiao2023} (with the Einasto profile of index 0.43), the optical depth toward the LMC center is $4.2\times 10^{-7}$. For the \citet{ou2024} model (Einasto profile of index 1.1), that is referenced by \citet{hawkins2025}, the optical depth toward the LMC center is $4.1 \times 10^{-7}$. These values should be compared to the optical depth of $4.4 \times 10^{-7}$ found in our fiducial model \citep{cautun2020}. Taken at face value, even if either the \citet{jiao2023} or \citet{ou2024} Milky Way halo models were correct, the expected number of microlensing events by hypothetical compact objects would only differ from the calculations presented by \citet{mroz2024b} by a few percent at most.

Parenthetically, \citet{hawkins2025} refer to their earlier paper \citep{bellido2024}, in which they claim that they reanalyzed the MACHO data using a newer Milky Way halo model by \citet{jiao2023}. Unfortunately, we were unable to reproduce the optical depth values shown in their Fig.~2. Specifically, \citet{bellido2024} claim that the optical depth toward a source star located in the LMC center is $\sim 2 \times 10^{-7}$, whereas our calculations (described below) indicate that it is a factor of two larger ($4.2\times 10^{-7}$).

\begin{figure}[htb]
\centering
\includegraphics[width=.7\textwidth]{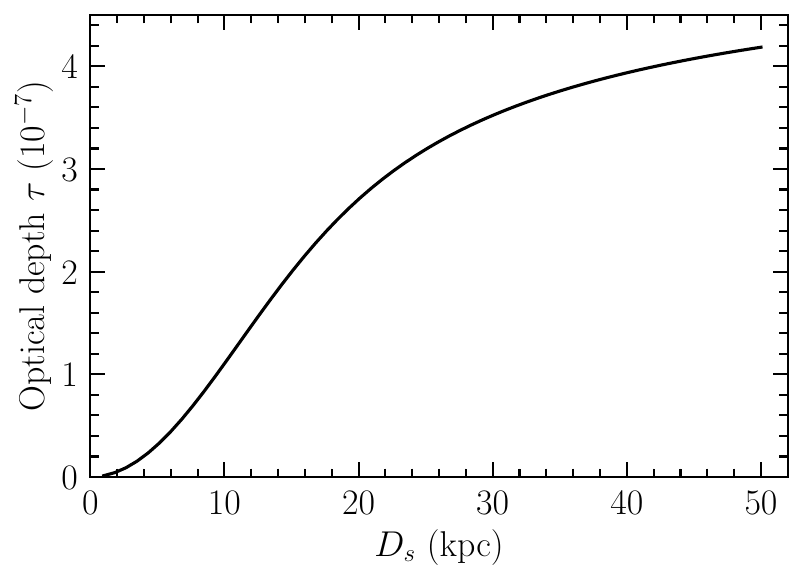}
\FigCap{Microlensing optical depth toward the LMC center as a function of source's distance assuming the Milky Way halo model of \citet{jiao2023}.}
\label{fig:tau}
\end{figure}

In an attempt to reproduce Fig.~2 of \citet{bellido2024}, we calculated the microlensing optical depth for a grid of source distances $0 < D_s < 50$\,kpc, using Eq.~(\ref{eq:tau}). The dark matter density in model B2 of \citet{jiao2023} is given by the formula
\begin{equation}
    \rho(R) = \rho_0 \exp\left[-\left(\frac{R}{R_h}\right)^{1/n}\right],
\end{equation}
where $R$ is the Galactocentric distance, $\rho_0=0.01992\,M_{\odot}\,\mathrm{pc}^{-3}$, $R_h=11.41$\,kpc, and $n=0.43$. For a lens located at a distance $D_l$ from Earth at the Galactic coordinates $(l,b)$, its Galactocentric distance is calculated as follows:
\begin{equation}
    R=\sqrt{D_l^2+R_0^2-2D_l R_0 \cos l \cos b},
\end{equation}
where $R_0=8.178$\,kpc is the distance to the Galactic center \citep{gravity2019}. Fig.~\ref{fig:tau} shows the optical depth as a function of the source distance in the direction of the LMC center ($l=280.82^{\circ}$, $b=-33.14^{\circ}$), which does not match that shown in \citet{bellido2024} at all. Because their optical depths are incorrect, we cannot place trust in the other calculations presented in that paper.

The best method to deduce the Milky Way rotation curve is by using tracers that allow for the calculation of complete 6D phase-space information. Recently, \citet{medina2025} combined the radial velocity data from the Dark Energy Spectroscopic Instrument (DESI) and proper motions from \textit{Gaia} with precise distances to RR Lyrae and blue horizontal-branch stars. They successfully calculated precise distances and 3D velocity vectors for 110 RR Lyrae stars and 330 blue horizontal-branch stars, all located at Galactocentric distances ranging from 50 to 100 kpc. Then, using this 6D information, they inferred the mass profile of the Milky Way. They found the total mass of the Milky Way within 50 kpc equal to $(0.38 \pm 0.05) \times 10^{12}\,M_{\odot}$ using RR Lyrae stars, and $(0.38 \pm 0.03) \times 10^{12}\,M_{\odot}$ using blue horizontal-branch stars (with the error bars representing 75\% confidence intervals), a factor of two more than in the \citet{jiao2023} and \citet{ou2024} models. These mass estimates translate to a circular velocity of 180\,km\,s$^{-1}$ at the LMC distance. Notably, these numbers are close to those obtained using the \citet{cautun2020} model. Thus, the latest data from DESI and \textit{Gaia} support the accuracy of the Milky Way halo model used by \citet{mroz2024b}.

\section{Final Conclusions}
\label{sec:conclusions}

Since the start of the first microlensing surveys in the 1990s, considerable progress has been made in developing techniques for extracting photometry in dense stellar regions. Many of the issues raised by \citet{hawkins2025} relate to how crowding affects the quality of photometric measurements. While these issues could have posed challenges for the early microlensing experiments that used PSF fitting photometry, the introduction of the DIA method in the 2000s largely eliminated these problems.

If the entire Milky Way dark matter halo were made of primordial black holes, whose mass function is described by the Thermal History Model by \citet{carr2021b,carr2024}, the combined OGLE-III and OGLE-IV experiments should have detected over 500 microlensing events \citep{mroz2024b}. However, only 13 events were actually detected \citep{mroz2024a,mroz2024b}. \citet{hawkins2025} tried to argue that OGLE is somehow missing a large fraction of expected events, perhaps missing those that exhibit finite-source effects or those with variable sources. As discussed in detail in Section~\ref{sec:selection}, the arguments presented by \citet{hawkins2025} lack a solid basis or are simply incorrect.

\citet{hawkins2025} also argued that the event detection efficiencies calculated by \citet{mroz2024a} are incorrect. However, as discussed in Section~\ref{sec:efficiency}, simulations used to calculate these detection efficiencies are consistent with the observed properties of the detected events. Moreover, the methodology for calculating detection efficiencies adopted by \citet{mroz2019,mroz2024a} has been validated by independent measurements of the microlensing optical depth toward the Galactic bulge by the MOA collaboration \citep{nunota2025}.

Many comments brought forward by \citet{hawkins2025} are related to the discrepancy between the results obtained by the OGLE survey and those published by the MACHO project \citep{alcock2000b}. As demonstrated in Sections~\ref{sec:macho} and \ref{sec:dia}, at least five event candidates identified by \citet{alcock2000b} have variable sources. Two of these objects exhibit repeating outbursts and are classified as members of a newly identified group of cataclysmic variables, dubbed as millinovae \citep{mroz2024c}. Removing these five variable objects from the sample of \citet{alcock2000b} reduces the MACHO measurement of the microlensing optical depth toward the LMC by almost 40\%.

OGLE is not the only microlensing experiment that did not confirm the MACHO findings. Only one candidate event was found by the EROS survey \citep{tisserand2007} from a sample of 7 million stars observed over more than 6.7 years (from 1996 to 2003). This led them to calculate a 95\% upper limit on the microlensing optical depth toward the LMC of $\tau < 0.36 \times 10^{-7}$, much lower than that found by \citet{alcock2000b} ($1.2^{+0.4}_{-0.3} \times 10^{-7}$). Additionally, no convincing long-timescale microlensing events were discovered in the combined MACHO and EROS data \citep{moniez2022}.

The results of the OGLE and EROS studies are further supported by various other pieces of evidence. For example, observations of supermagnified stars in lensing clusters demonstrate that compact objects with masses above $10^{-6}\,M_{\odot}$ may account for at most 2\% of dark matter \citep{muller2025}. A similar study by \citet{kawai2024} found that no more than 20\% of the dark matter in the direction of the supermagnified star Icarus may be made of PBHs of $\sim 1\,M_{\odot}$.
\citet{shah2025}, using observations of 1532 Type Ia supernovae collected by the Dark Energy Survey, concluded that at most 12\% of dark matter over cosmological distances may be made of compact objects with masses greater than $0.03\,M_{\odot}$. Moreover, the discovery of a sizable population of wide-separation binary stars in the ultrafaint dwarf galaxy Bo\"otes\,I indicates that compact objects with masses larger than $5\,M_{\odot}$ cannot account for more than 1\% of the dark matter in that galaxy \citep{shariat2025}.

Overall, these studies collectively indicate that no large population of compact objects (including PBHs) with masses greater than $10^{-8}\,M_{\odot}$ exists in dark matter.

{\it And yet they are not found} -- this is the conclusion of the most extensive ever OGLE microlensing survey of the Magellanic Clouds, as well as other surveys including MACHO and EROS.

\section*{Acknowledgments}

We thank Ilaria Caiazzo for her advice on the paper's title. We thank all the OGLE observers for their contribution to the collection of the photometric data over the decades. This research was funded in part by National Science Centre, Poland, grants OPUS 2021/41/B/ST9/00252 and SONATA 2023/51/D/ST9/00187 awarded to P.M. The OGLE project has received funding from the Polish National Science Centre grant OPUS 2024/55/B/ST9/00447 awarded to A.U. {\L}.W. acknowledges support from the Polish National Science Centre DAINA grant No 2024/52/L/ST9/00210.

\bibliographystyle{acta}
\bibliography{pap}

\section*{Appendix A}

\begin{figure}
\includegraphics[width=\textwidth]{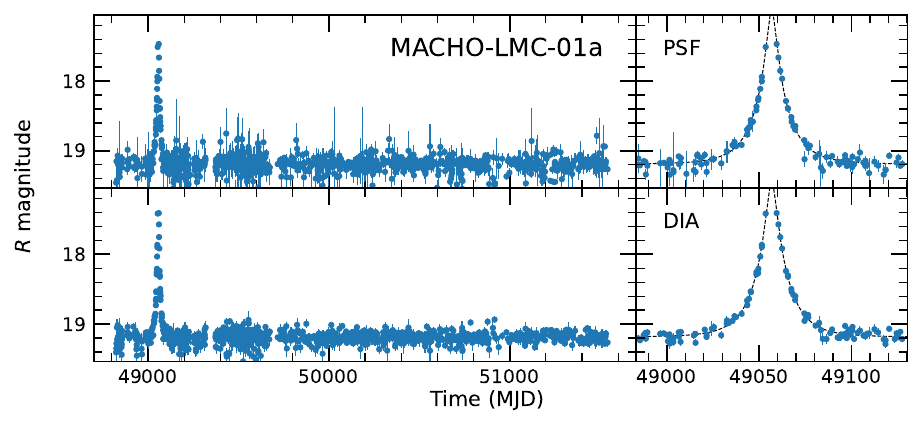}
\includegraphics[width=\textwidth]{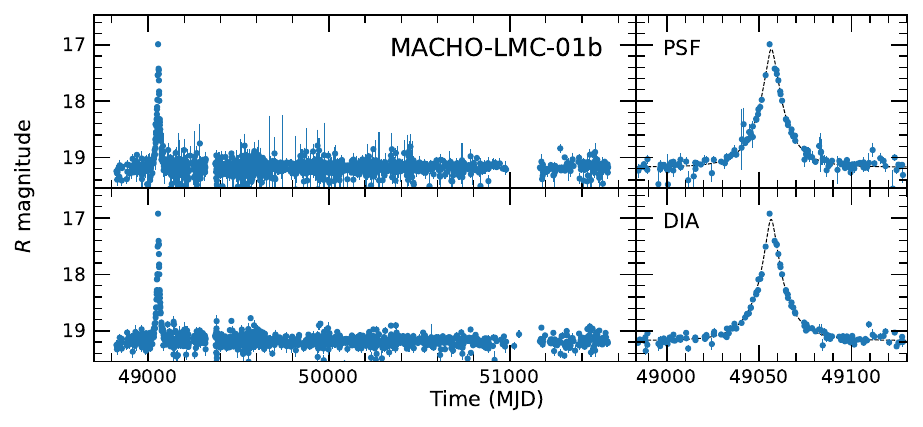}
\includegraphics[width=\textwidth]{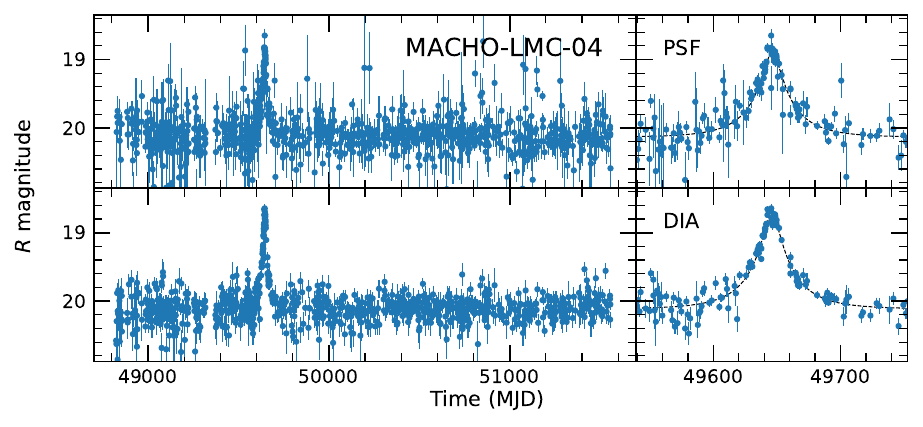}
\caption{Comparison between PSF and DIA $R$-band light curves of MACHO microlensing candidates.}
\label{fig:app1}
\end{figure}

Figs.~\ref{fig:app1} to \ref{fig:app2} present the comparison between the original PSF and new DIA $R$-band light curves of microlensing event candidates identified by \citet{alcock2000b}.

\newpage

\begin{figure}
\includegraphics[width=\textwidth]{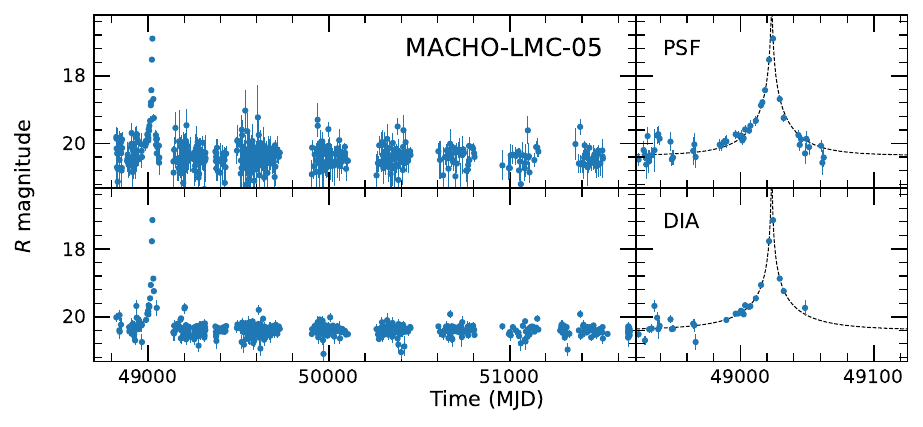}
\includegraphics[width=\textwidth]{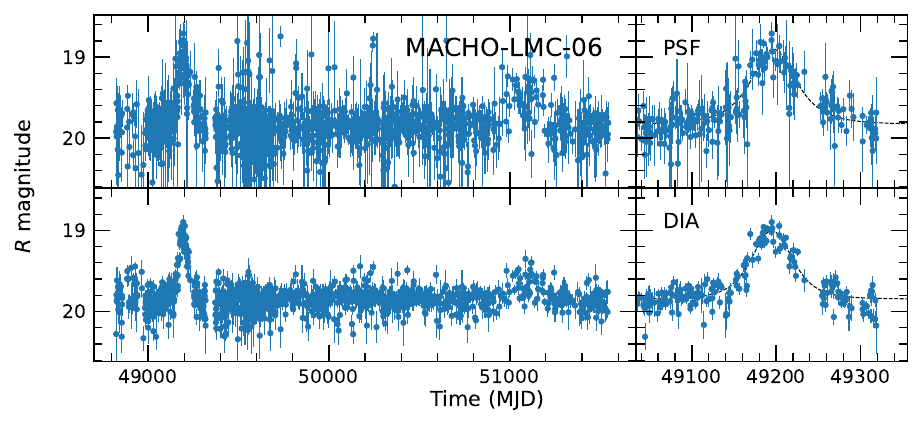}
\includegraphics[width=\textwidth]{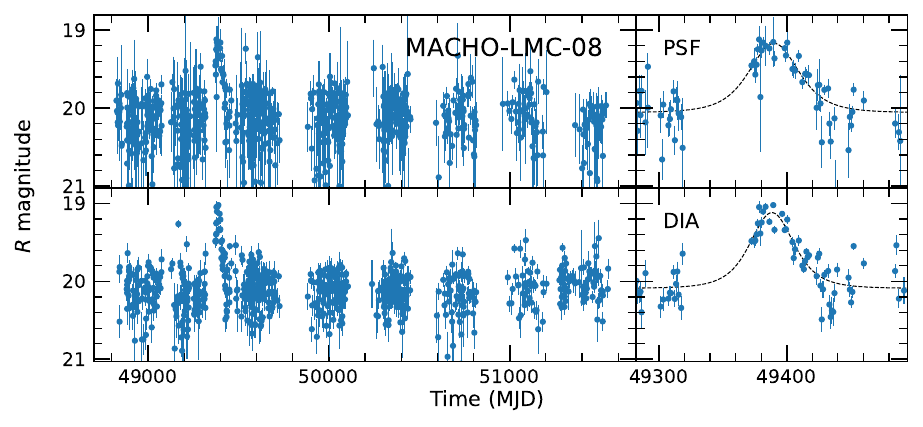}
\caption{Comparison between PSF and DIA $R$-band light curves of MACHO microlensing candidates.}
\end{figure}

\begin{figure}
\includegraphics[width=\textwidth]{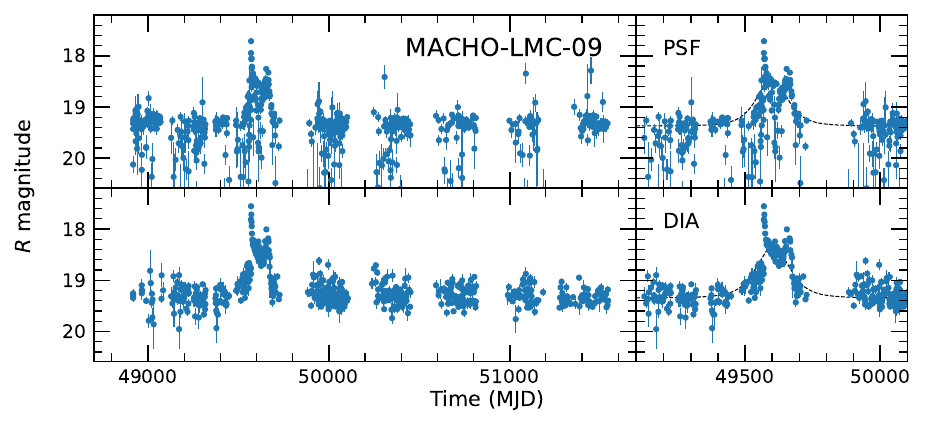}
\includegraphics[width=\textwidth]{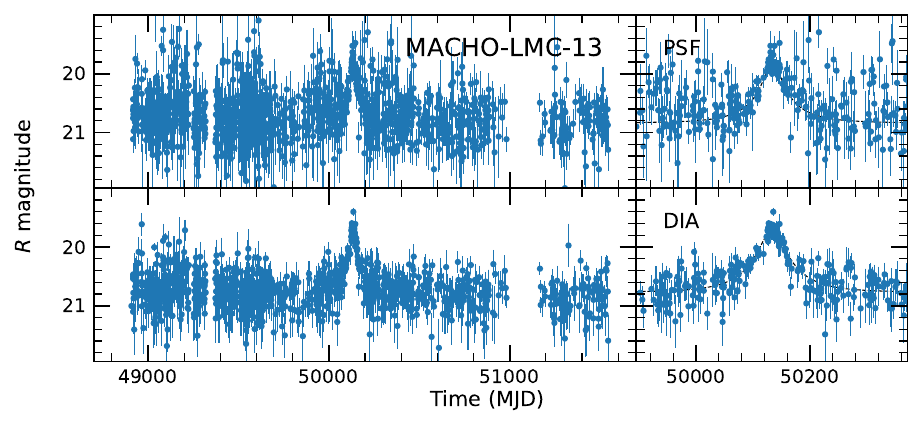}
\includegraphics[width=\textwidth]{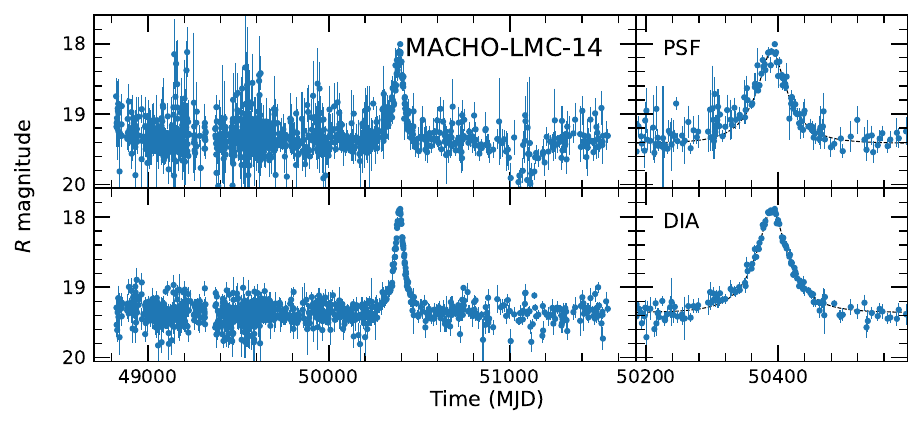}
\caption{Comparison between PSF and DIA $R$-band light curves of MACHO microlensing candidates.}
\end{figure}

\begin{figure}
\includegraphics[width=\textwidth]{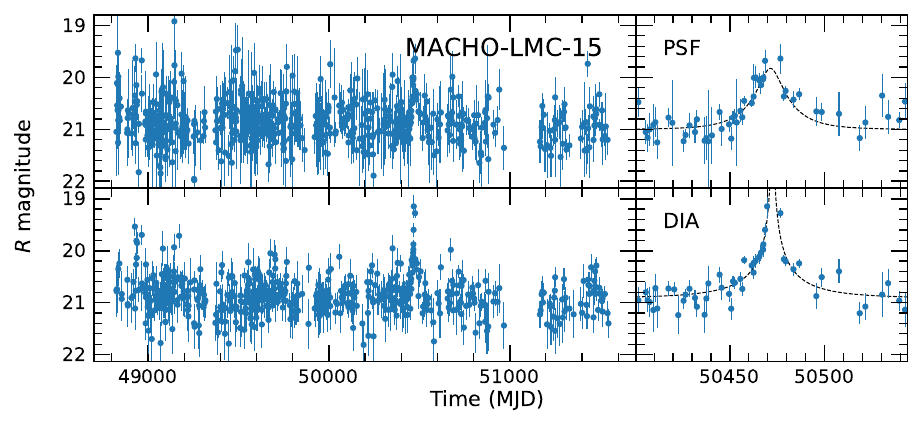}
\includegraphics[width=\textwidth]{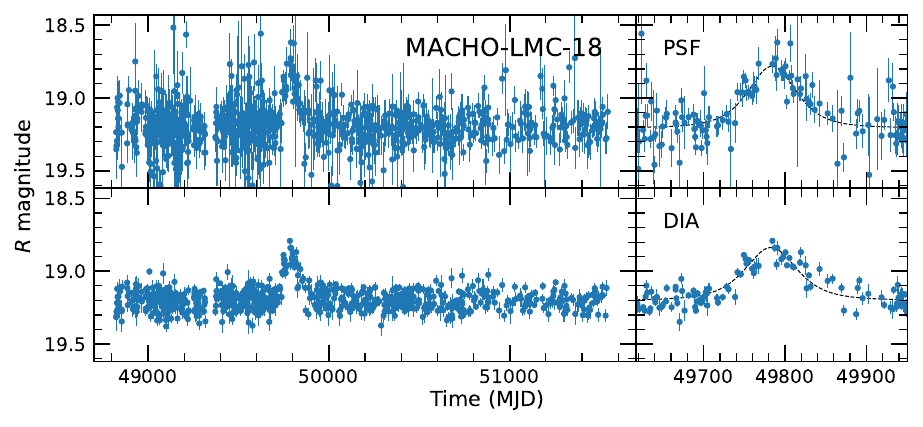}
\includegraphics[width=\textwidth]{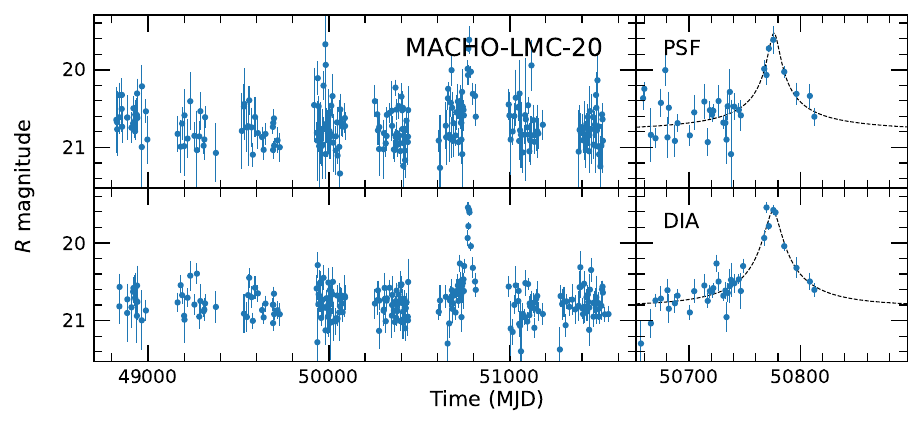}
\caption{Comparison between PSF and DIA $R$-band light curves of MACHO microlensing candidates.}
\end{figure}

\begin{figure}
\includegraphics[width=\textwidth]{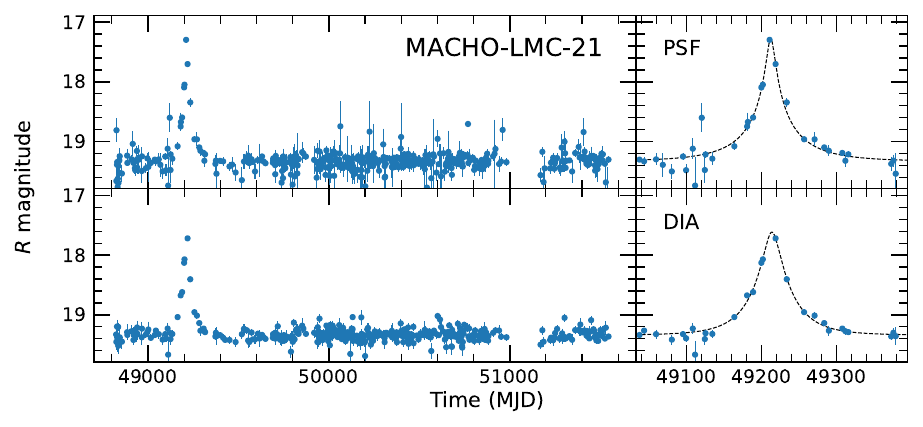}
\includegraphics[width=\textwidth]{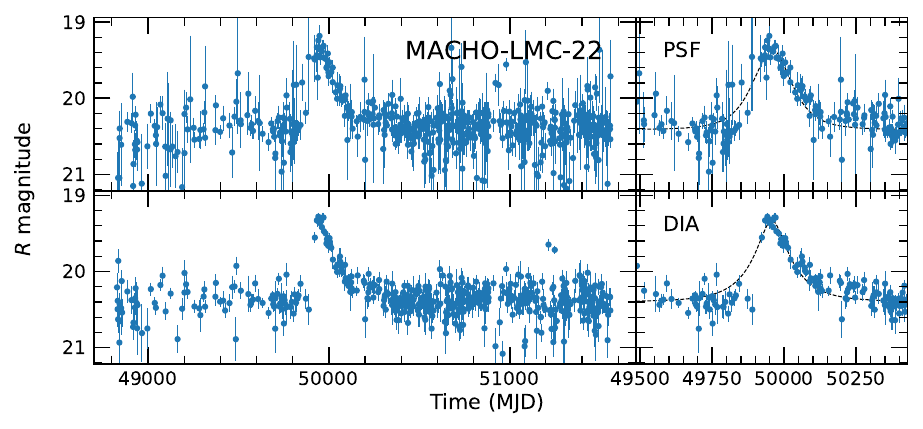}
\includegraphics[width=\textwidth]{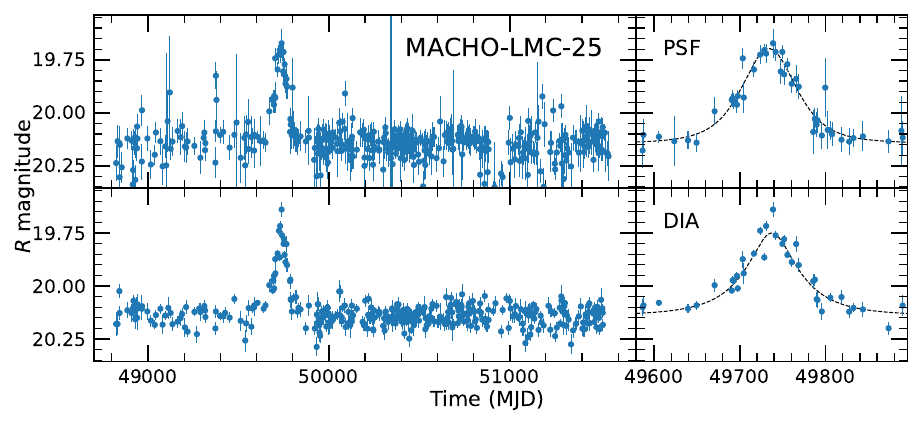}
\caption{Comparison between PSF and DIA $R$-band light curves of MACHO microlensing candidates.}
\end{figure}

\begin{figure}
\includegraphics[width=\textwidth]{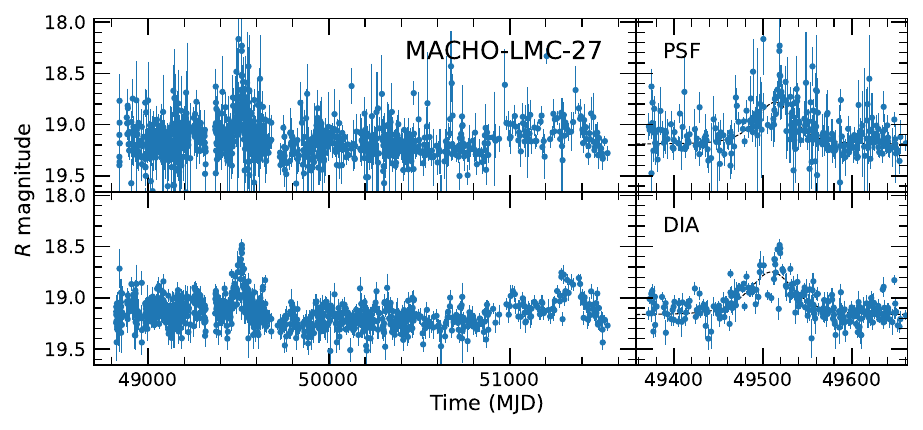}
\includegraphics[width=\textwidth]{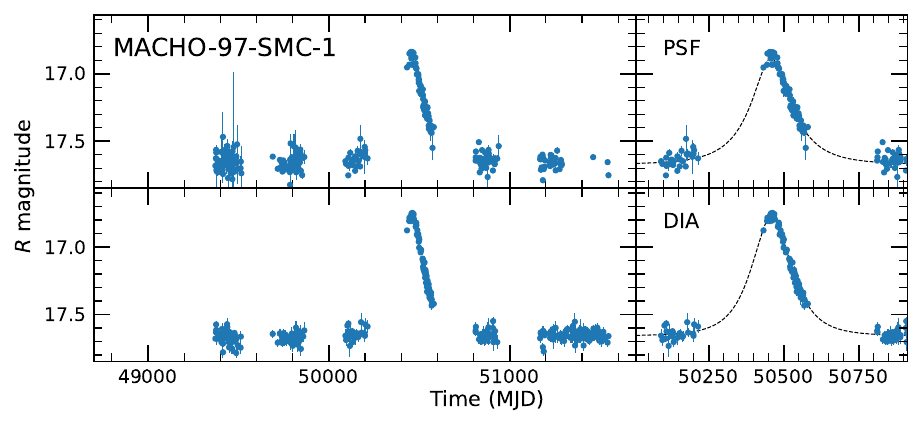}
\caption{Comparison between PSF and DIA $R$-band light curves of MACHO microlensing candidates.}
\label{fig:app2}
\end{figure}

\end{document}